\documentclass[twocolumn,tighten,times]{aastex62}

\usepackage{mathrsfs}
\usepackage{amsmath}

\newcommand{\hb}{\ifmmode {\rm H\beta} \else H$\beta$\fi}
\newcommand{\mgii}{\ifmmode {\rm Mg\ II} \else Mg {\sc ii}\fi}
\newcommand{\feii}{\ifmmode {\rm Fe\ II} \else Fe {\sc ii}\fi}
\newcommand{\heii}{\ifmmode {\rm He\ II} \else He {\sc ii}\fi}
\newcommand{\oiii}{\ifmmode {\rm [O\ III]} \else [O {\sc iii}]\fi}
\newcommand{\rblr}{\ifmmode {R_{\rm BLR}} \else $R_{\rm BLR}$\fi}
\newcommand{\rhb}{\ifmmode {R_{\rm H\beta}} \else $R_{\rm H\beta}$\fi}
\newcommand{\taublr}{\ifmmode {\tau_{\rm BLR}} \else $\tau_{\rm BLR}$\fi}
\newcommand{\mbh}{\ifmmode {M_{\bullet}} \else $M_{\bullet}$\fi}
\newcommand{\fblr}{\ifmmode {f_{\rm BLR}} \else $f_{\rm BLR}$\fi}
\newcommand{\rl}{$R_{\rm H\beta}$--$L_{5100}$}
\newcommand{\dotm}{\ifmmode {\dot{\mathscr{M}}} \else $\dot{\mathscr{M}}$\fi}
\newcommand{\fwhmhb}{\ifmmode {\rm FWHM_{\hb}} \else $\rm FWHM_{\hb}$\fi}
\newcommand{\Rfe}{\ifmmode {{\cal R}_{\rm Fe}} \else ${\cal R}_{\rm Fe}$\fi}
\newcommand{\Dhb}{\ifmmode {{\cal D}_{\hb}} \else ${\cal D}_{\hb}$\fi}
\newcommand{\ewoiii}{\ifmmode {{\rm EW}_{\rm [OIII]}} \else ${\rm EW}_{\rm [OIII]}$\fi}
\newcommand{\ewheii}{\ifmmode {{\rm EW}_{\rm HeII}} \else ${\rm EW}_{\rm HeII}$\fi}
\newcommand{\ewhb}{\ifmmode {{\rm EW}_{\rm H\beta}} \else ${\rm EW}_{\rm H\beta}$\fi}

\begin{document}

\title{\large \bf The Radius-Luminosity Relationship Depends on Optical 
Spectra in Active Galactic Nuclei}

\author{Pu Du}
\affiliation{Key Laboratory for Particle Astrophysics, Institute of High Energy Physics,
Chinese Academy of Sciences, 19B Yuquan Road, Beijing 100049, China}

\author{Jian-Min Wang}
\affiliation{Key Laboratory for Particle Astrophysics, Institute of High Energy Physics,
Chinese Academy of Sciences, 19B Yuquan Road, Beijing 100049, China}
\affiliation{National Astronomical Observatories of China, Chinese Academy of Sciences, 20A Datun Road, Beijing 100020, China}
\affiliation{School of Astronomy and Space Science, University of Chinese Academy of Sciences, 19A Yuquan Road, Beijing 100049, China}

\begin{abstract} 
The radius-luminosity (\rl) relationship of active galactic nuclei (AGNs)
established by the reverberation mapping (RM) observations has been widely used
as a single-epoch black hole mass estimator in the research of large AGN
samples. However, the recent RM campaigns discovered that the AGNs with high
accretion rates show shorter time lags by factors of a few comparing with the
predictions from the \rl\ relationship. The explanation of the shortened time
lags has not been finalized yet. We collect 8 different single-epoch spectral
properties to investigate how the shortening of the time lags correlate with
those properties and to understand what is the origin of the shortened lags. We
find that the flux ratio between \feii\ and \hb\ emission lines shows the most
prominent correlation, thus confirm that accretion rate is the main driver for
the shortened lags. In addition, we establish a new scaling relation including
the relative strength of \feii\ emission. This new scaling relation can provide
less biased estimates of the black hole mass and accretion rate from the
single-epoch spectra of AGNs. 
\end{abstract}

\keywords{galaxies: active; galaxies: nuclei - quasars: supermassive black holes}

\section{Introduction}
\label{sec:intro}

In the past 40 years, reverberation mapping (RM; e.g., \citealt{bahcall1972,
blandford1982, peterson1993}) has become a powerful tool to investigate the
geometry and kinematics of the broad-line regions (BLRs) in active galactic
nuclei (AGNs) and to measure the masses of supermassive black holes (BHs).
Through long-term spectroscopic monitoring of an AGN, the size scale (\rblr) of
its BLR can be directly obtained by measuring the delayed response (\taublr) of
the emission line with respect to the variation of the continuum, where
$\rblr=c\taublr$ and $c$ is the speed of light. Fortunately, the RM observations of
$\sim100$ objects \citep[e.g.,][]{peterson1993, peterson1998, peterson2002,
peterson2004, kaspi2000, kaspi2007, bentz2008, bentz2009, denney2009, barth2011,
barth2013, barth2015, rafter2011, rafter2013, du2014, du2015, du2016V, du2018,
du2018b, wang2014, shen2016, fausnaugh2017, grier2012, grier2017, derosa2018, woo2019}
lead to a correlation between the time lag of \hb\ emission line (or the radius
\rhb\ of the \hb-emitting region) and the monochromatic luminosity ($\lambda
L_{\lambda}$) at 5100 \AA\ (hereafter $L_{5100}$) with the form of
\begin{equation}
\label{eqn:r-l_bentz}
\rhb = \alpha \ell^{\beta}_{44}, 
\end{equation}
where $\ell_{44} = L_{5100} / 10^{44}\ {\rm erg\ s^{-1}}$
\citep[e.g.,][]{kaspi2000, bentz2009b, bentz2013}. This correlation makes it
possible to estimate BLR radius from a single-epoch spectrum. It is
called the \rl\ relationship, and has been widely adopted as a single-epoch
BH mass estimator in the research of large AGN samples
\citep[e.g.,][]{mclure2004, vestergaard2006, kollmeier2006, greene2007,
shen2011}. However, there is growing evidence for increasing scatters of the 
\rl\ relationship from ongoing RM campaigns. 

\begin{figure*}
    \centering
    \includegraphics[width=0.48\textwidth]{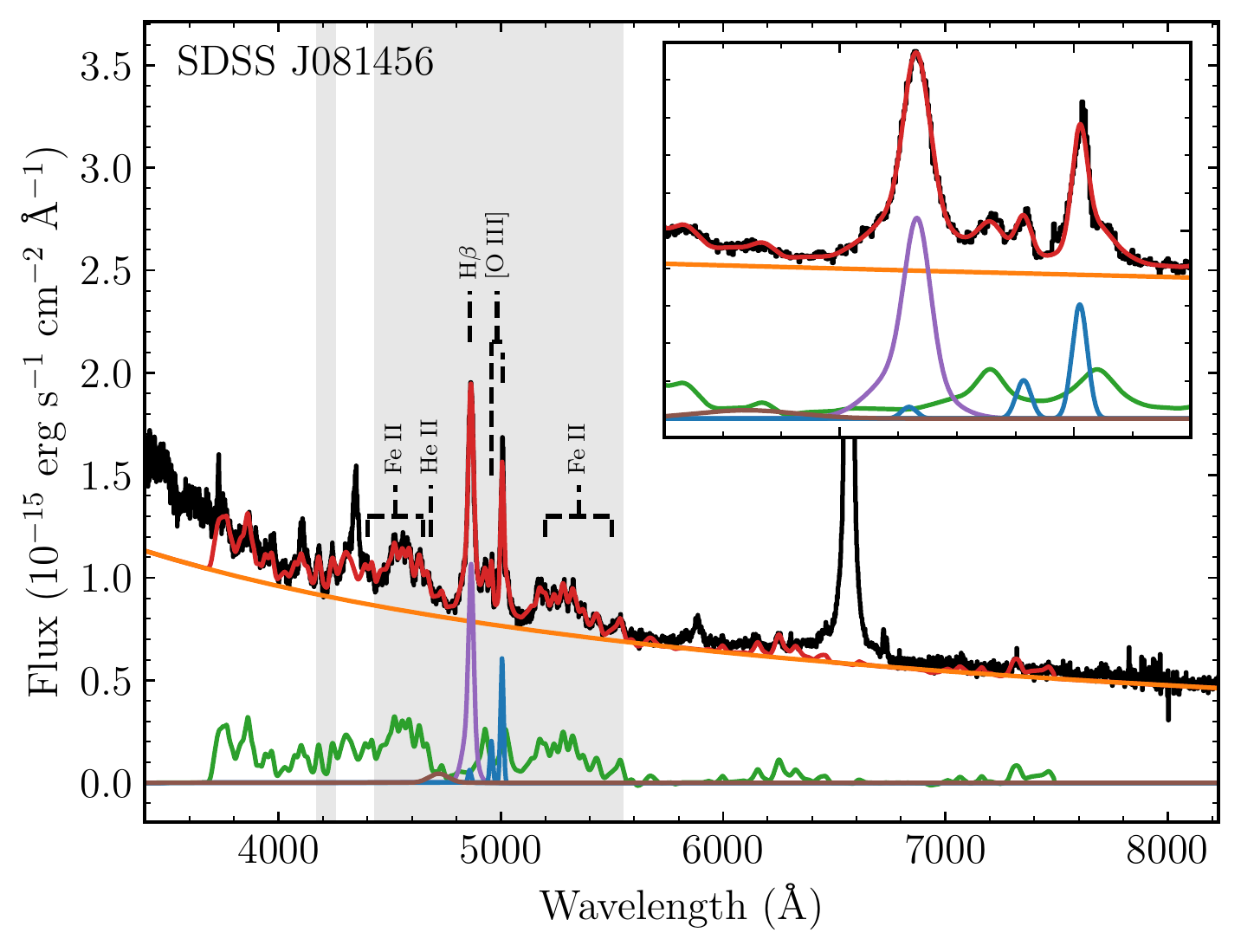}\hspace{0.3cm}\includegraphics[width=0.48\textwidth]{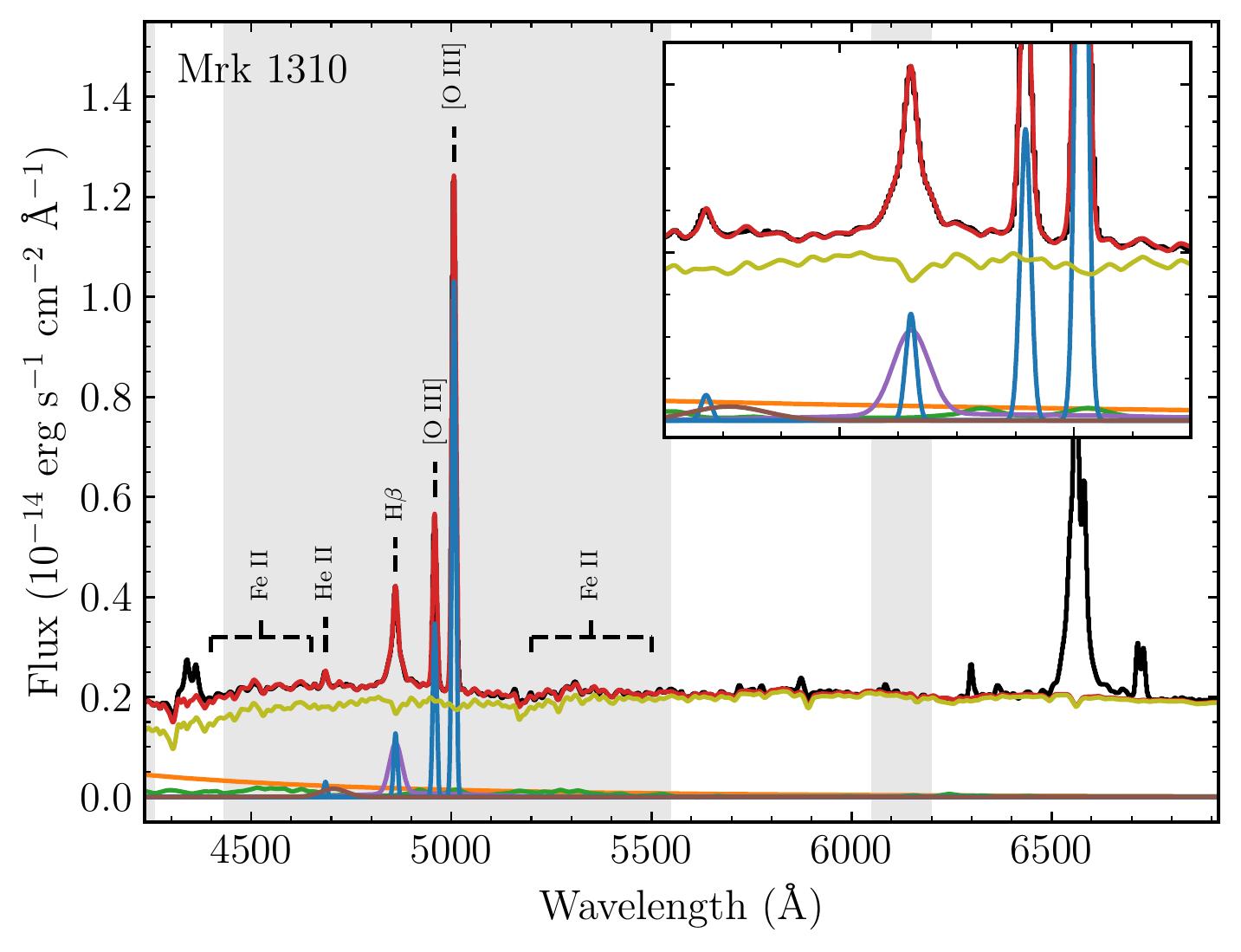}
    \caption{Two fitting examples. In each panel, the black line is the spectrum
    in the rest frame after Galactic extinction correction. The red line is the
    best fit. The orange line is the power law of the continuum. The purple line
    is the broad \hb\ component. The blue lines are the narrow emission lines
    (\oiii$\lambda\lambda4959,5007$, \hb, and \heii). The green, brown, and
    yellow lines are the \feii\ template, broad \heii, and the template of host
    galaxy, respectively. The zoom-in panels show the detailed fitting around
    \hb. Some strong emission lines are labeled.}
    \label{fig:fitting}
\end{figure*}

Reverberation of broad emission lines to the continuum confirms photoionization
as the major radiation mechanism in the BLR. As a canonic model of BLR,
photoionization defined by the ionization parameter $\Xi \propto L_{\rm ion} /
R_{\rm BLR}^2 n_{\rm e} T$ directly indicates $\rblr \propto L_{\rm ion}^{1/2}$,
where $L_{\rm ion}$ is the ionizing luminosity, $n_{\rm e}$ and $T$ are electron
density and temperature of the ionized gas, respectively. If we take $L_{5100}$
as a proxy of $L_{\rm ion}$, we have $\rblr \propto L_{5100}^{1/2}$, agreeing
with the observations \cite[see also][]{bentz2013}. We would like to emphasize
here the necessary conditions for this canonic relation: (1) ionizing source
should be isotropic or at least quasi-isotropic so that the BLR clouds
receive the same luminosity with observers; (2) $L_{\rm ion} \propto L_{5100}$
should always work; (3) ionizing luminosity comes from a point source, which is
much smaller than the distances of the BLR clouds to the central BH. Condition
(1) is broken in the AGNs powered by slim accretion disks \citep{wang2014c},
where the puffed-up inner region may lead to non-isotropic
ionizing radiation \citep{wang2014c}. Condition (2) relies on spectral energy
distributions (only holds for pro-grade accretion AGNs powered by the
Shakura-Sunyaev disks), and does not work in ones with retro-grade accretion
\citep{wang2014b, czerny2019}. Low spin (also low accretion rate and large
BH mass) may lead to the deficit of the UV photons and non-linear relation
between $L_{\rm ion}$ and $L_{5100}$ \citep{wang2014b, czerny2019}. And the
$L_{5100}$ variation shows a little lag with respect to the $L_{\rm ion}$
variation in the RM of accretion disks \citep[e.g.,]{edelson2015, mchardy2018,
cackett2018}. About Condition (3), the size of accretion disk, although
small, has been successfully measured and is not infinitesimal
\citep{edelson2015, mchardy2018, cackett2018}. Therefore, the \rl\
relationship is expected to depend on accretion situation or some other
properties. 

Recently, Super-Eddington Accreting Massive Black Hole (SEAMBH) campaign
discovered that many objects with strong \feii\ and narrow \hb\ emission lines,
which are thought to be the AGNs with high accretion rates, lie below the \rl\
relationship \citep{du2015, du2016V, du2018}. They found that the time lags of
the AGNs with high accretion rates become shortened by factors of $3\sim8$
relative to the normal-accretion-rate AGNs with the same luminosities, and the
shortening itself shows correlation with the accretion rate \citep{du2015,
du2016V, du2018}. \cite{wang2014c} proposed that the anisotropic radiation of
the slim accretion disk may probably result in the shortened time lags. The
Sloan Digital Sky Survey Reverberation Mapping (SDSS-RM) Project also reported
many AGNs have time lags shorter than expected from the \rl\ relationship
\citep{grier2017}, but cautioned that selection effects may arise at least in
some cases \citep[see][]{grier2019}.

Although the detailed physical explanation causing the shortened time lags is
not yet finalized \citep{wang2014c, grier2017, grier2019}, more and more objects
deviating from the traditional \rl\ relationship are being discovered
\citep{grier2017, du2018}. It is urgent to investigate the origin of the
shortened lags in more detail. In this paper, we investigate how the deviation
of an AGN from the \rl\ relationship correlates with the properties in the
single-epoch spectrum, and try to establish a new scaling relationship including
the influence of single-epoch spectral properties. We describe the sample, data,
and measurements in Section \ref{sec:data_measurement}. A new scaling relation
is established and presented in Section \ref{sec:analysis}. Some discussions are
provided in Section \ref{sec:discussions}, and a brief summary is given in
Section \ref{sec:summary}. We adopt the standard $\Lambda$CDM cosmology and the
parameters of $H_0=67~{\rm km~s^{-1}~Mpc^{-1}}$, $\Omega_{\Lambda}=0.68$, and
$\Omega_m=0.32$ \citep{ade2014, planck2018} in this paper.

\section{Data and Measurement}
\label{sec:data_measurement}

\subsection{Sample}

The analysis in the present paper is mainly based on the samples: (1) the RM
measurements compiled in \cite{bentz2013} from the previous literatures, (2) the
AGNs with high accretion rates of the SEAMBH campaign published in \cite{du2014,
du2015, du2016V, du2018, wang2014, hu2015}, (3) some other AGNs published after
2013: Mrk~1511 from \cite{barth2013}, NGC~5273 from \cite{bentz2014},
KA~1858+4850 from \citep{pei2014}, MCG~+06-30-015 from \citep{bentz2016b,
hu2016}, UGC~06728 from \citep{bentz2016a}, and MCG~+08-11-011, NGC~2617, 3C~382
and Mrk~374 from \cite{fausnaugh2017}\footnote{We also include the new RM
observations of the previous mapped objects after 2013 in the following
analysis: NGC~4593 \citep{barth2013}, NGC~7469 \citep{peterson2014}, NGC~5548
\citep{lu2016, pei2017}, NGC~4051 \citep{fausnaugh2017}, PG~1226+023 (3C~273,
\citealt{zhang2019}), and PG~2130+099 \citep{hu2019}}. The collection in
\cite{bentz2013} includes 41 AGNs monitored successively since the late 1980's,
most of which have relatively weaker \feii\ emission and broader \hb\ lines
compared to the SEAMBH objects \citep{du2018}. The SEAMBH campaign, as a
dedicated RM project for AGNs with high accretion rates, published the time lags
of 25 objects (totally 30 measurements, some objects have more than one
measurement). Including the objects published after 2013, we have totally 75
objects with 117 measurements. The time lags, 5100\AA, \hb, and \oiii\
luminosities, equivalent width (EW) of \hb\ and \oiii, and \hb\ FWHM of the
corresponding campaigns are listed in Table \ref{tab:rm_results}. Some objects
have been mapped more than once, in order to understand the population
properties better, we have to equalize the weights of the individual objects in
the following analysis. We average the multiple measurements by taking into
account their measurement uncertainties \cite[see more details in][]{du2015}.
The average measurements, etc. lags, FWHM, luminosities, are also listed in
Table \ref{tab:rm_results}.

\begin{figure*}
    \centering
    \includegraphics[width=\textwidth]{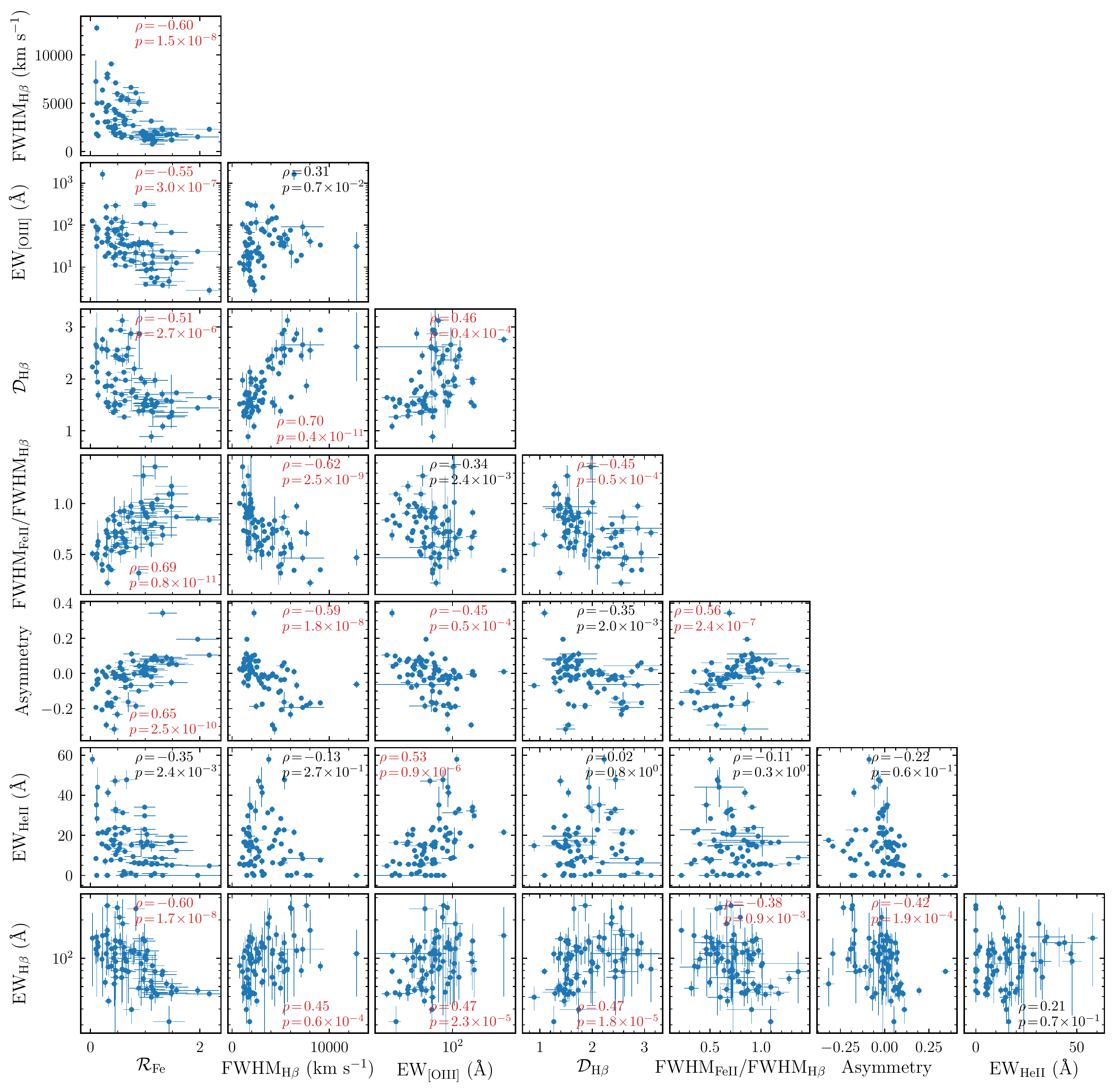}
    \caption{Pairwise correlations of the single-epoch spectral properties 
    for the sample in the present paper. 
    The Spearman's rank correlation coefficients ($\rho$) and the corresponding
    null probabilities (see Section \ref{sec:pairwise}) are marked in the 
    corner of each panel (with red color if $p < 0.001$). }
    \label{fig:correlations}
\end{figure*}

\subsection{Data of Single-epoch Spectral Properties}
\label{sec:spec_properties}

To investigate how the single-epoch spectral properties control the deviation of
AGNs from the \rl\ relationship, we compile 8 different parameters from the
spectra around the \hb\ region in the optical band: (1) the flux ratio between
\feii\ and \hb, which is denoted as 
\begin{equation}
\Rfe=F_{\rm Fe}/F_{\hb},
\end{equation}
where $F_{\rm Fe}$ is the flux of Fe {\sc ii} from 4434\AA\ to 4684\AA\ and
$F_{\hb}$ is the flux of broad \hb, (2) FWHM of \hb\ emission line (\fwhmhb),
(3) equivalent width (EW) of \oiii$\lambda5007$ emission line (\ewoiii), (4) the
ratio between FWHM and $\sigma_{\rm line}$ (second moment of the line profile)
of the \hb\ line ($\Dhb=\fwhmhb/\sigma_{\rm \hb}$), (5) the ratio between
the FWHM of \feii\ and the FWHM of \hb\ (${\rm FWHM}_{\feii}/{\rm FWHM}_{\hb}$),
(6) the asymmetry of \hb\ line defined by
$A=[\lambda_c(3/4)-\lambda_c(1/4)]/\fwhmhb$, where $\lambda_c(3/4)$ and
$\lambda_c(1/4)$ are the central wavelengths at the 3/4 and 1/4 of the \hb\ peak
height \citep{derobertis1985, boroson1992, brotherton1996, du2018b}, (7) the EW
of \heii\ (\ewheii), and (8) the EW of \hb\ (\ewhb). These parameters are
referred to as ``single-epoch spectral properties'', because they can be
measured simply from the single-epoch spectra rather than from the time-domain
observations like RM\footnote{Of course, if we have the RM data, we can
definitely measure them from an individual spectrum in the RM campaign or the
mean spectrum (can be treated as the average of the values from the individual
spectra). But if we don't have the RM data, we can still measure them from the
single-epoch spectra found in some other literatures or databases.}. In order to
establish some new scaling relationships which can be applied to the large AGN
samples obtained in the spectroscopic surveys of SDSS or Dark Energy
Spectroscopic Instrument (DESI) in the near future, we need to find the
correlations between the single-epoch properties and the deviation of AGNs from
the \rl\ relationship. The FWHM and EW of the \hb\ line, and the EW of the
\oiii\ line of each RM campaign are collected and listed in Table
\ref{tab:rm_results}. We use them (and the average values for the objects with
multiple RM measurements) directly in the following analysis. We search the
other parameters in the literatures and list the values in Table
\ref{tab:properties}. For the parameters that we can not find in the
literatures, we fit the spectra of the objects found in the public archive and
measure those parameters by ourselves (see also Table \ref{tab:properties}).

\begin{figure*}
    \centering
    \includegraphics[width=0.79\textwidth]{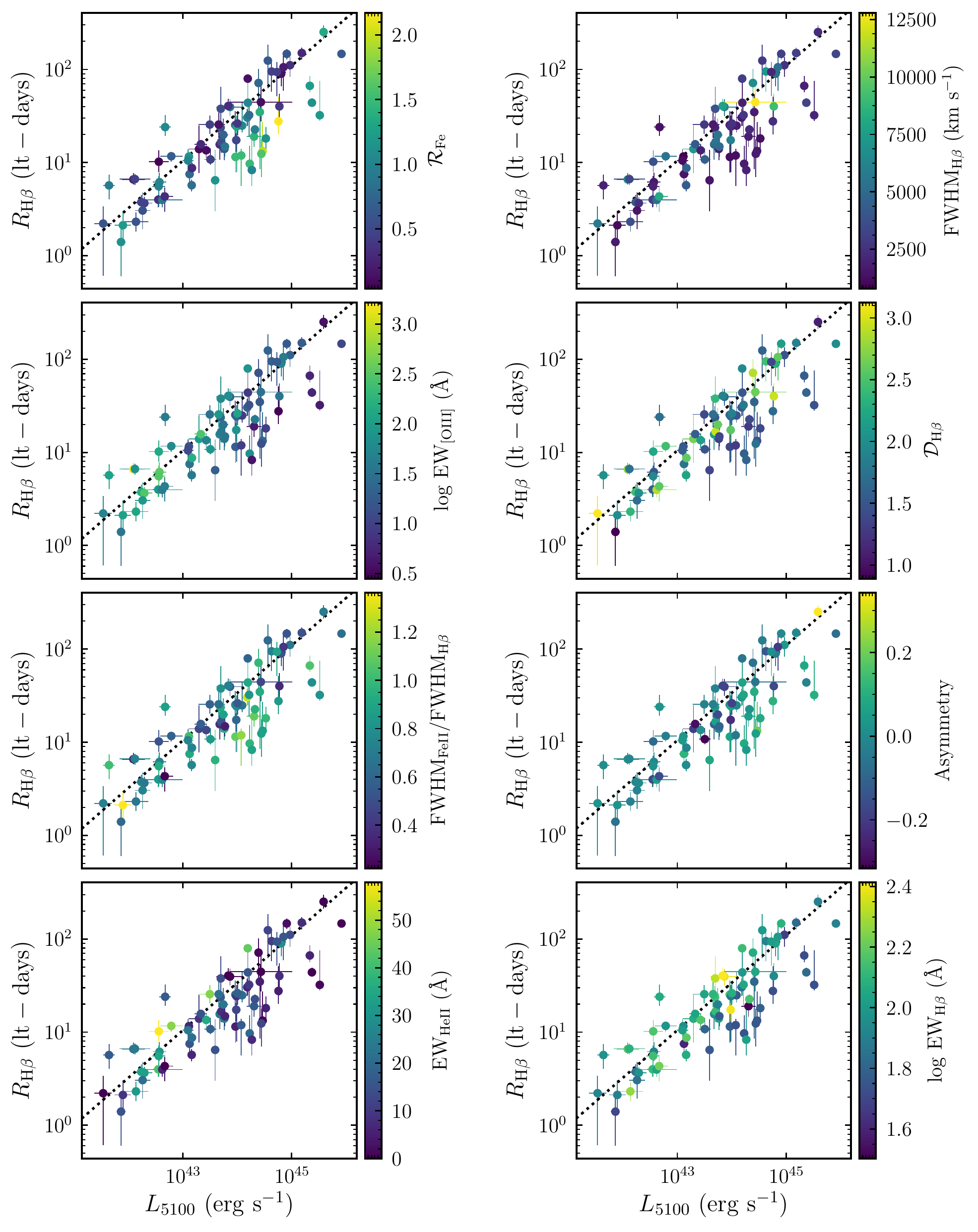}
    \caption{The \rl\ relationship color-coded by the spectral properties. The
    dotted lines are the \rl\ relationship for the low-accretion-rate AGNs in
    \cite{du2018}. It is obvious that some objects deviate from the dotted
    lines. The colors show clear trend with \Rfe\ and \ewhb, which means the
    deviation correlates with \Rfe\ and \ewhb.}
    \label{fig:r_l}
\end{figure*}

\begin{figure*}
    \centering
    \includegraphics[width=\textwidth]{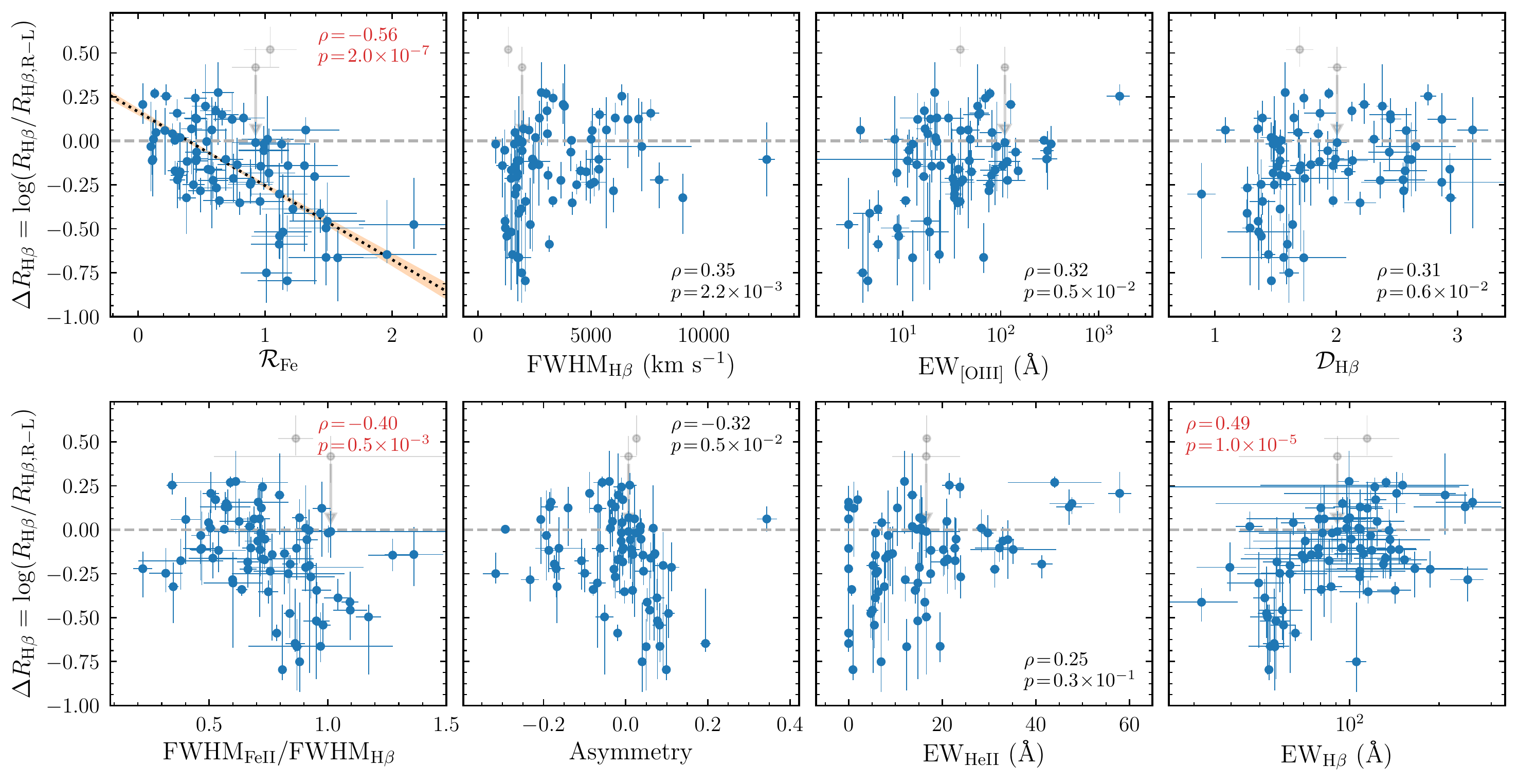}
    \caption{The correlations between $\Delta \rhb$ and the properties
    for the sample in the present paper. 
    The Spearman's rank correlation coefficients ($\rho$) and the corresponding
    null probabilities (see Section \ref{sec:pairwise}) are marked in the 
    corner of each panel (with red color if $p < 0.001$). \Rfe\ 
    shows significant correlation with $\Delta \rhb$, and \ewhb\ shows 
    a moderate correlation. The grey dashed lines show $\Delta \rhb=0$.
    The black dotted line and the orange color show the linear regression 
    and its confidence band (2$\sigma$). The two grey points 
    are MCG~+06-26-012 and MCG~+06-30-015 (see more details in 
    Section \ref{sec:new_scaling_relation}). We correct the intrinsic reddening
    MCG~+06-30-015 and use its corrected luminosity in the analysis. 
    We do not include MCG~+06-26-012 in the analysis.}
    \label{fig:deltaR}
\end{figure*}

\subsection{Fitting the spectra}
\label{sec:fitting}

We use the following components in the spectral fitting: (1) a power law to
model the AGN continuum, (2) two Gaussians to model the broad \hb\ emission
line, (3) a template constructed from the \feii\ spectrum of the narrow-line
Seyfert 1 (NLS1) galaxy I Zw 1 by \cite{boroson1992} for the \feii\ emission,
(4) one or two Gaussians for each of the narrow emission lines, e.g.,
\oiii$\lambda\lambda$4959,5007, \hb, \heii\ (if necessary), (5) one or two
Gaussians to model the broad \heii\, and (6) a simple stellar population
model\footnote{We select the simple stellar population model which can get the
smallest $\chi^2$ in the fitting.} from \cite{bruzual2003} as a template for the
contribution of the host galaxy if necessary. The fitting is mainly performed in
the windows of 4170--4260\AA\ and 4430--5550\AA\ in the rest frame. If the host
contribution is significant, we supplement the window of 6050--6200\AA\ to give
a better constraint to the fitting of stellar template. All of the narrow-line
components in each object are fixed to have the same velocity width and shift,
except for those showing very different width/shift. \oiii$\lambda$4959 is fixed
to have one-third of the \oiii$\lambda$5007 flux \citep{osterbrock2006}. NLS1s
always show very weak narrow emission lines (in particular, the narrow \hb). In
the fitting for the spectra of NLS1s, we also fix the flux of narrow \hb\ to be
one-tenth of the \oiii$\lambda$5007 flux \citep{kewley2006, stern2013}. Two
examples (SDSS~J081456 and Mrk~1310) of the multi-component spectral fitting are
shown in Figure \ref{fig:fitting}. The contribution of host galaxy in the
spectrum of SDSS~J081456 is weak, while Mrk~1310 is host-dominated. The fitting
of these two objects are fairly good. We measure the spectral properties (see
Section \ref{sec:spec_properties}) from the fitting results and list them in
Table \ref{tab:properties}.

\section{Analysis}
\label{sec:analysis}

\subsection{Pairwise Correlations between Different Properties}
\label{sec:pairwise}

Before discussing the correlations between the single-epoch spectral properties
and the deviation from the \rl\ relationship, we first present the pairwise
correlations between different properties in Figure \ref{fig:correlations}.
Although there are many similar discussions using different samples in the
historical literatures \citep[e.g.,][]{boroson1992}, it is still valuable to do
this demonstration for the RM objects. Spearman's rank correlation coefficients
($\rho$) and the corresponding two-sided $p$-value for a null hypothesis test
(two sets of data are uncorrelated) are marked in the panels of Figure
\ref{fig:correlations}. The $\rho$ and $p$ values of the significant
correlations (with $p<0.001$) are marked with red color.

Among all of the correlations, \fwhmhb\ versus \Dhb\ is the most significant
one, which has Spearman's correlation coefficient $\rho=0.70$. It means that if
the width of \hb\ line is smaller, its profile tends to be more Lorentzian-like.
This correlation has been demonstrated by, e.g., \cite{kollatschny2011,
kollatschny2013}. The sample in the present paper is larger than that used in
\cite{kollatschny2011, kollatschny2013}, but the result is almost the same. 
The
secondarily-significant correlations are \Rfe\ versus ${\rm FWHM}_{\rm
FeII}/\fwhmhb$ ($\rho=0.69$), \Rfe\ versus \hb\ Asymmetry ($\rho=0.65$), \Rfe\
versus \ewhb\ ($\rho=-0.60$), \Rfe\ versus \fwhmhb\ ($\rho=-0.60$), and \fwhmhb\
versus ${\rm FWHM}_{\feii}/\fwhmhb$ ($\rho=-0.62$). The previous three
correlations mean that, if the relative strength of \feii\ is higher, the widths
of \feii\ and \hb\ are more similar, \hb\ line tends to have stronger blue wing,
and the EW of \hb\ line is weaker. Similar to the correlation between \Rfe\ and
${\rm FWHM}_{\feii}/\fwhmhb$, \fwhmhb\ versus ${\rm FWHM}_{\feii}/\fwhmhb$
means that the objects with narrower \hb\ lines also have more similar \feii\
and \hb\ widths. In addition, the correlations of \Rfe\ versus \ewoiii, \Rfe\
versus \Dhb, \fwhmhb\ versus \hb\ Asymmetry, \ewoiii\ versus \ewheii, and ${\rm
FWHM}_{\feii}/\fwhmhb$ versus \hb\ Asymmetry are also significant. \Rfe\
versus (\fwhmhb\ and \ewoiii) are the prominent correlations in the famous AGN
eigenvector 1 sequence \citep[e.g.,][]{boroson1992, sulentic2000, marziani2001,
marziani2003, marziani2018, shen2014, sun2015}. The detailed physical process of this
sequence is still under some debate \citep[e.g.,][]{panda2019}. The correlation
between \Rfe\ versus \hb\ Asymmetry has been demonstrated using the PG quasar
sample in \cite{boroson1992}, and is also associated with the eigenvector 1
sequence. The RM sample reproduces this correlation. Besides, there are some
other weak correlations, please see Figure \ref{fig:correlations}. More
discussions about the pairwise correlations are provided in Section
\ref{sec:discussions}.

\subsection{Deviation from the \rl\ Relationship}
\label{sec:deviation}

In Figure \ref{fig:r_l}, we first show the \rl\ relationship color-coded by the
properties we collected. The deviation from the \rl\ relationship shows clear
correlation with \Rfe\ and \ewhb, both of which
show obvious variation trend across the \rhb\ axis in Figure \ref{fig:r_l}. In
order to further investigate the significance of the correlations, we define the
deviation from the \rl\ relationship as
\begin{equation}
    \Delta \rhb = \log\, (\rhb/R_{\rm H\beta, R-L}),
\end{equation}
where $R_{\rm H\beta, R-L}$ is the prediction from the \rl\ relationship. Here,
we adopt $\log R_{\rm H\beta, R-L}=1.53+0.51\log \ell_{44}$ obtained by
\cite{du2018} for the AGN with dimensionless accretion rate $\dotm<3$ 
(\dotm\ is defined by the following Equation (\ref{eqn:mdot}), please see 
Section \ref{sec:accretion_rate})
as the fiducial \rl\ relationship.
It should be noted that using the \rl\ relationship in \cite{bentz2013} doesn't
change the discussion and conclusion in this paper. 

The correlations between $\Delta \rhb$ and the single-epoch properties are shown
in Figure \ref{fig:deltaR}. Again, the Spearman's correlation coefficients and
the corresponding null probabilities are marked in the corners of the panels in
Figure \ref{fig:deltaR}. The correlation between $\Delta \rhb$ and \Rfe\ is the
most significant one. $\Delta \rhb$ shows a strong anti-correlation with \Rfe\
with Spearman's correlation coefficient $\rho = -0.56$ and the null probability
$p = 2\times10^{-7}$. $\Delta \rhb$ and \ewhb\ shows a weaker correlation with
$\rho = 0.49$ and $p = 1\times10^{-5}$. In addition, the low-\Dhb\ or
small-\fwhmhb\ objects show more extended distribution of $\Delta \rhb$, while
the high-\Dhb\ or large-\fwhmhb\ objects have relatively narrower $\Delta \rhb$
distribution and the average $\Delta \rhb$ more close to zero. But the
Spearman's coefficients of $\Delta \rhb$ versus \Dhb\ and \fwhmhb\ are not high
enough. 
Actually, this complex distribution in the $\Delta \rhb$ versus \fwhmhb\
(or \Dhb) plane may be caused by the eigenvector 1 sequence (the correlation
between \Rfe\ and \fwhmhb, see, e.g., \citealt{boroson1992, sulentic2000, marziani2001,
marziani2003, shen2014, sun2015}, or the recent review in
\citealt{marziani2018}). The large-\fwhmhb\ (or
high-\Dhb) objects have low \Rfe\ values, but the small-\fwhmhb\ (or low-\Dhb)
objects have large range of \Rfe\ (span from low to high \Rfe). Therefore, the
$\Delta \rhb$ distribution in the small-\fwhmhb\ (or low-\Dhb) objects is more
extended.
Some more discussions about this are provided in Section
\ref{sec:orientation}. The correlations between $\Delta \rhb$ and the other
parameters are not significant given the present data ($\Delta \rhb$ shows weak 
a correlation with ${\rm FWHM_{\feii}}/\fwhmhb$). Because the correlation
coefficient of \Rfe\ is the highest among all of the spectral properties, \Rfe\
can be regarded as the primary parameter that controls the deviation of an AGN
from the \rl\ relationship. 

The eigenvector 1 sequence (or main sequence) of AGNs has been extensively
investigated in the past decades, and contains the information of the evolution
or systematic variation of AGNs (see the recent review in
\citealt{marziani2018}). Through the analysis of the eigenvector 1 sequence, 
\Rfe\ has been demonstrated as a probe of accretion rate/Eddington ratio (e.g.,
\citealt{boroson1992, sulentic2000, marziani2001, marziani2003, shen2014,
sun2015}), thus a primary physical driver of the shortened time lags is the
accretion rate. 

As a simple test, we provide here the linear regression of the correlation
between $\Delta \rhb$ and \Rfe. We adopt the BCES method \citep[][the orthogonal
least squares]{akritas1996} to perform the linear regression, which takes into
account both of the error bars in x and y axis. MCG~+06-26-012 has a relatively
low sampling cadence in the first 80 days in its light curve of \cite{wang2014} and
\cite{hu2015}, which makes its time lag may bias towards longer value. We do not use
it in the regression. And the intrinsic reddening of MCG~+06-30-015 is strong in
light of its high Balmer decrement \citep{hu2016}. We correct its intrinsic
reddening and use the corrected luminosity and the corresponding $R_{\rm H\beta, R-L}$. The MCG~+06-26-012 and
MCG~+06-30-015 are marked as grey points in Figure \ref{fig:deltaR} (also in the
following Figure \ref{fig:new_r_l}). The linear regression is yielded as
\begin{equation}
    \Delta \rhb = -(0.42\pm0.06) \Rfe  + (0.17\pm0.05).
    \label{eqn:DeltaR_Rfe}
\end{equation}
The regression and the corresponding confidence band ($2\sigma$) are shown in
Figure \ref{fig:deltaR}. We have also tested that the residual $\Delta \rhb -
\Delta \rhb (\Rfe)$ does not show any correlations with all of the spectral
properties (with Spearman's coefficients $|\rho|<0.25$), where $\Delta \rhb
(\Rfe)$ is the $\Delta \rhb$ value deduced from \Rfe\ by Equation
(\ref{eqn:DeltaR_Rfe}).

\begin{figure}
    \centering
    \includegraphics[width=0.47\textwidth]{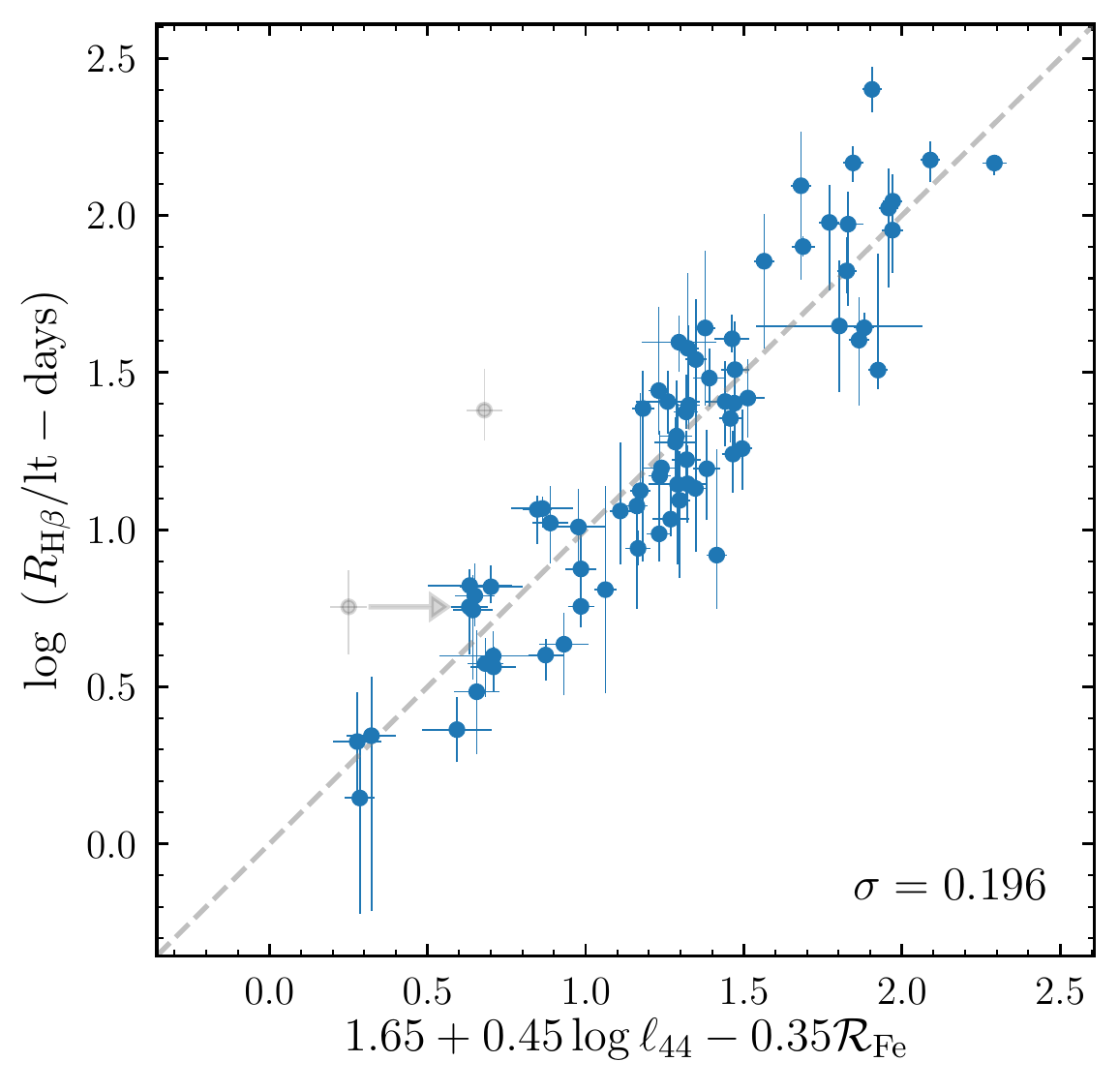}
    \caption{New scaling relation. The scatter of the new scaling relation
    is $\sigma=0.196$, and marked in the lower right corner. The two grey points 
    are MCG~+06-26-012 and MCG~+06-30-015 (see more details in 
    Section \ref{sec:new_scaling_relation}). We correct the intrinsic reddening
    MCG~+06-30-015 and use its corrected luminosity in the analysis. 
    We do not include MCG~+06-26-012 in the analysis.}
    \label{fig:new_r_l}
\end{figure}

\subsection{A New Scaling Relation}
\label{sec:new_scaling_relation}

Because the strongest correlation is the relation between $\Delta \rhb$ and
\Rfe, we can add \Rfe\ as a new parameter into the \rl\ relationship to
establish a new scaling relation with smaller scatter. We fit the RM sample with
the following new scaling relation:
\begin{equation}
    \label{eqn:new_scaling_relation}
    \log\, (\rhb/{\rm lt\!-\!days}) = \alpha + \beta \log \ell_{44} + \gamma \Rfe.
\end{equation}
In order to obtain the uncertainties of the parameters, we employ the bootstrap
technique. A subset is generated by resampling $N$ points from the RM sample
with replacement ($N$ is the number of the objects in the RM sample). Then, we
calculate the best parameters for this subset using the Levenberg-Marquardt
method \citep{press1992}, and repeat this procedures for 5000 times to generate
the distributions of $\alpha$, $\beta$, and $\gamma$. The final best parameters
and the corresponding uncertainties are obtained from the $\alpha$, $\beta$, and
$\gamma$ distributions.  The fit is shown in Figure \ref{fig:new_r_l}, and the best
parameters are:
\begin{equation}
    \alpha = 1.65\pm0.06, \beta = 0.45\pm0.03, \gamma = -0.35\pm0.08.
\end{equation}
The scatter of the new scaling relation is $\sigma=0.196$, which is much smaller
than the original scatter of the \rl\ relationship \citep[$\sigma\sim0.28$,
see][]{du2018}.

\section{Discussions}
\label{sec:discussions}

\subsection{Some Discussions on Pairwise Correlations}
\label{sec:discussion_pairwise}

In Section \ref{sec:pairwise}, we showed the pairwise correlations between the
parameters we compiled. Some of the correlations have been presented in the
literatures using different samples of AGNs, and some have been discussed
directly or indirectly. A $\sigma_{\hb}$-\Dhb\ correlation, which is a
width-profile correlation of the \hb\ line similar to the \fwhmhb-\Dhb\ in this
paper, was presented in \cite{collin2006} using the RM sample at that time. It
was also discussed by \cite{kollatschny2011, kollatschny2013} and explained as
the different contributions from the rotation/Keplerian motions and the turbulent
velocities in the objects with different line widths \citep{kollatschny2011,
kollatschny2013}. The correlation in Figure \ref{fig:correlations} is generally
the same as in \cite{kollatschny2011}, but has more objects at the narrow-width
(small \fwhmhb) end because the current sample has more NLS1s or
high-accretion-rate objects. However, it should be noted that the parameter
\Dhb\ involves \fwhmhb\ as the numerator, thus may 
introduce a certain degree of self correlation to the \fwhmhb-\Dhb\ relation. 

The comparison between ${\rm FWHM_{\feii}}$ and \fwhmhb\ has been presented for
the quasar sample in the Sloan Digital Sky Survey (SDSS) in, e.g.,
\cite{hu2008a, hu2008b, cracco2016}. The ${\rm FWHM_{\feii}}$ is systematically
smaller than \fwhmhb, which was demonstrated and explained by the contribution
from a intermediate-line region in \cite{hu2008a, hu2008b}. The \feii\ emission
and the intermediate-width component of \hb\ line are both from this
intermediate-line region, and \hb\ has an extra very broad component
\citep{hu2008a, hu2008b}. In addition, \cite{hu2015} shows a comparison between
the time lags of \feii\ and \hb\ using the SEAMBH sample, which is also a direct
evidence for the relatively larger size of \feii-emitting region and the smaller
\hb-emitting region. And \cite{hu2015} also shows that the lag ratio between
\feii\ and \hb\ correlates with \Rfe. The $\Rfe-{\rm FWHM_{\feii}}/\fwhmhb$
correlation in Figure \ref{fig:correlations} may has the same physical origin as
the correlation between the lag ratio and \Rfe\ (The gas in the \feii\ region
has a larger size and a smaller FWHM). Similarly, \fwhmhb\ correlates with ${\rm
FWHM_{\feii}}/\fwhmhb$ (see in Figure \ref{fig:correlations}) because of the
\Rfe-\fwhmhb\ correlation \citep[Eigenvector 1 sequence, e.g.,][]{boroson1992,
sulentic2000, marziani2001, marziani2003, marziani2018, shen2014}.

The principal component analysis in \cite{boroson1992} has shown a weak
correlation between \Rfe\ and \ewhb\ using the PG quasar sample, however with a
relatively small correlation coefficient of $-0.425$ \citep[see more details
in][]{boroson1992}. The \Rfe\ and \ewhb\ of the RM sample presented in this
paper show a slightly stronger correlation with Spearman's coefficient of
$\rho=-0.60$. This correlation may be related to the Baldwin effect of \hb\ line
\citep[e.g.,][]{baldwin1977, korista2004}, and especially with the intrinsic
Baldwin effect \citep[e.g.,][]{gilbert2003, rakic2017}. The intrinsic Baldwin
effect shows an anti-correlation between EW of the emission line and the
luminosity, and is also equivalent to an anti-correlation between EW and the
accretion rate because the BH mass keeps a constant during the observation
campaign \citep[e.g.,][]{gilbert2003, rakic2017}. The \Rfe\ parameter is
correlated with Eddington ratio/accretion rate, thus is 
naturally correlated with the \ewhb.

\begin{figure}
    \centering
    \includegraphics[width=0.47\textwidth]{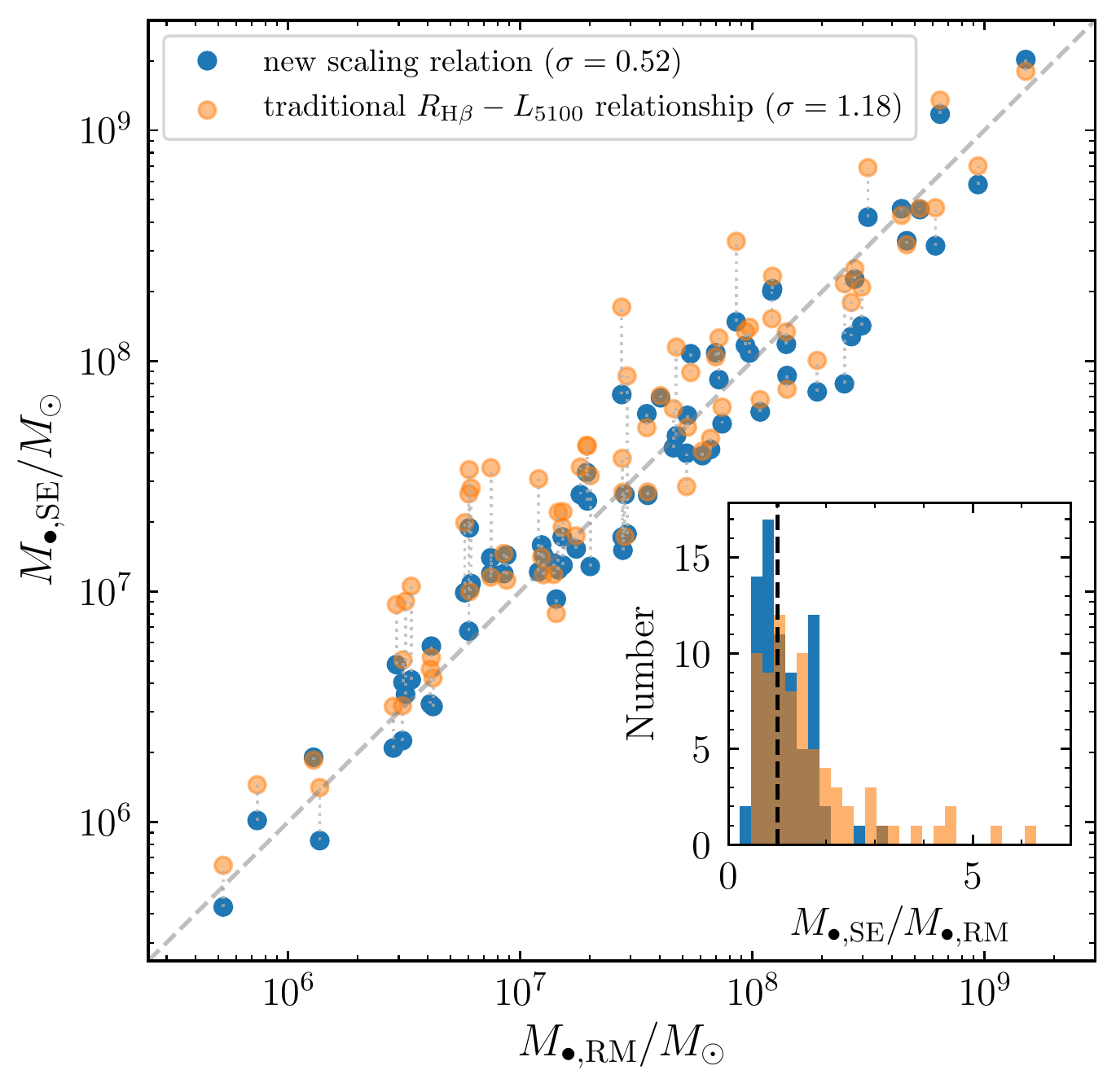}
    \caption{Single-epoch BH mass versus RM BH mass. The blue points are
    estimated by the new scaling relation, while the orange points are obtained
    by the traditional \rl\ relationship. The embedded panel shows the
    distributions of $M_{\bullet,{\rm SE}}/M_{\bullet,{\rm RM}}$ of the new
    scaling relation and the \rl\ relationship. The BH masses estimated by the
    \rl\ relationship are biased towards to higher values with respect to those
    from the new scaling relation. The standard deviations (simply denoted by
    $\sigma$) of the distributions are provided in the upper-left corner. We do
    not plot the error bars in order to show the differences more clearly.}
    \label{fig:mass_mass}
\end{figure}

\begin{figure*}
    \centering
    \includegraphics[width=\textwidth]{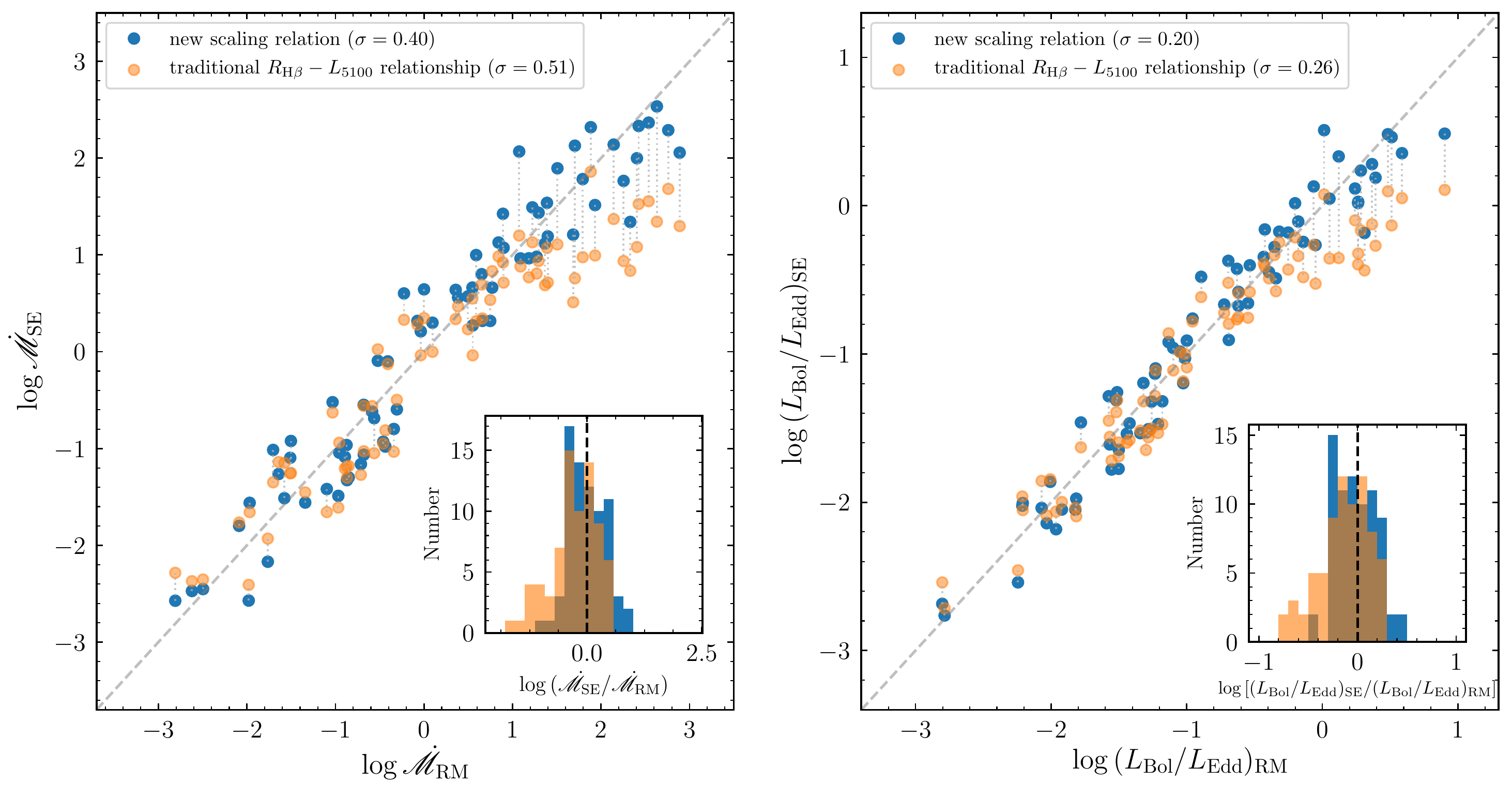}
    \caption{Accretion rates and Eddington ratios from the single-epoch spectra
    and the RM observations. The blue points are deduced by the new
    scaling relation, while the orange points are obtained by the 
    \rl\ relationship. The embedded panel shows the corresponding distributions
    of $\log\,(\dotm_{\rm SE}/\dotm_{\rm RM})$ and $\log\,[(L_{\rm Bol}/L_{\rm
    Edd})_{\rm SE}/(L_{\rm Bol}/L_{\rm Edd})_{\rm RM}]$. The accretion rates and
    Eddington ratios estimated from the \rl\ relationship are biased downward in
    the high-\dotm\ end. The standard deviations (simply denoted by $\sigma$) of
    the distributions are provided in the upper-left corner. We do not plot the
    error bars in order to show the differences more clearly.}
    \label{fig:mdot_mdot}
\end{figure*}

The asymmetry-\Rfe\ correlation has been shown in \cite{boroson1992}, and
discussed in the context of eigenvector 1 sequence \citep{sulentic2002}. The
high-\Rfe\ objects tend to have stronger blue \hb\ wings, and vise versa. The
${\rm FWHM_{\feii}}/\fwhmhb$-asymmetry and \fwhmhb-asymmetry correlations can
also attribute to the asymmetry-\Rfe\ correlation. The origin of the \hb\
asymmetry must be subject to the geometry and kinematics of the BLRs, but is
still under some debate because of the degeneracy of \hb\ profiles with
different BLR geometry and kinematics. A recent dedicated RM campaign project
for the BLR kinematics of the AGNs with \hb\ asymmetry\footnote{They may have
some special BLR kinematics or inhomogeneous gas distribution, or even binary
BHs in their centers (see more details in \citealt{du2018b}).} has started
\citep{du2018b}, and may provide more observations for the velocity-resolved RM
measurements in the future.

The \ewoiii-\ewheii\ correlation is a natural result of the photoionization
physics. Both of \oiii\ and \heii\ have high ionization energy (54.9 eV for
\oiii\ and 54.4 eV for \heii, respectively), thus are sensitive to the variation
of the AGN circumstances (e.g., spectral energy distribution, SED) in a similar way.

\begin{figure*}
    \centering
    \includegraphics[width=\textwidth]{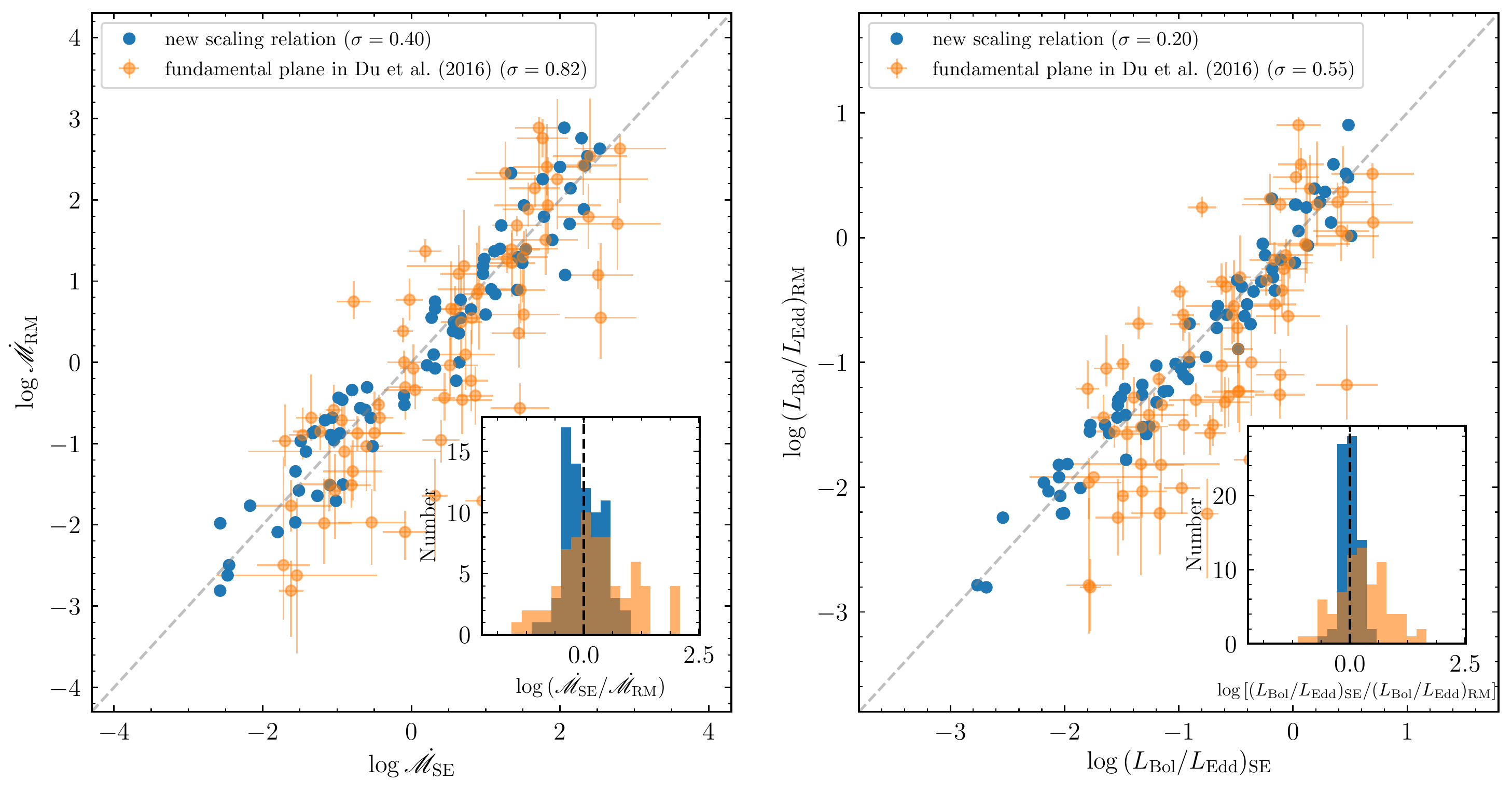}
    \caption{Comparisons with the accretion rates and Eddington ratios estimated
    from the fundamental plane of BLR. The blue points are deduced by the new
    scaling relation, while the orange points are obtained by the FP in
    \cite{du2016FP}. The embedded panel shows the corresponding distributions of
    $\log\,(\dotm_{\rm SE}/\dotm_{\rm RM})$ and $\log\,[(L_{\rm Bol}/L_{\rm
    Edd})_{\rm SE}/(L_{\rm Bol}/L_{\rm Edd})_{\rm RM}]$. The accretion rates and
    Eddington ratios estimated from the FP show much larger scatter. The
    standard deviations (simply denoted by $\sigma$) of the distributions are
    provided in the upper-left corner. We do not plot the error bars of the blue
    points for clarity.}
    \label{fig:mdot_mdot_fp}
\end{figure*}

\subsection{The New Scaling Relation and BH Mass Measurement}

Through the analysis in this paper, we found that the \Rfe\ parameter is the
dominant observational property in the scatter of the \rl\ relationship. 
This confirms the statement that the AGNs with high accretion rates
tend to have shortened lags in \cite{du2015, du2016V, du2018}. The shortened
time lags in high-\Rfe/high-accretion-rate AGNs imply smaller BLR scale sizes
and smaller BH mass estimates with respect to the traditional \rl\ relationship.

The \rl\ relationship was heavily utilized as a single-epoch BH mass estimator
in large AGN samples, and helped establish our paradigm for AGN evolution
\citep{mclure2004, vestergaard2006, kollmeier2006, greene2007, shen2011}. The
shortened time lags in high-\Rfe/high-accretion-rate AGNs make it vital to take
this into account in the BH mass estimation. Therefore, we suggest calculate the
BH masses from single-epoch spectra following the steps below: (1) obtain the
\rhb\ from the luminosities and the strength of \feii\ (\Rfe) using Equation
(\ref{eqn:new_scaling_relation}), (2) get the BH mass by the following Equation
(\ref{eqn:bh_mass}) from the line width and the estimated \rhb. Then, the
dimensionless accretion rate can be easily estimated by the following Equation
(\ref{eqn:mdot}) in Section \ref{sec:accretion_rate}.

In combination
with the velocity width ($\Delta V$) of the emission line, RM measurement yields
an estimate of BH mass as 
\begin{equation}
    \label{eqn:bh_mass}
    \mbh = \fblr \frac{\Delta V^2 \rblr}{G}
\end{equation}
\citep[e.g.,][]{wandel1999, peterson1999, peterson2004}, where $G$ is the
gravitational constant, and \fblr\ is the virial factor related to the geometry
and kinematics of the BLR \citep[e.g.,][]{onken2004, park2012, grier2013f,
ho2014, woo2015}. Although measuring BH mass through RM technique is feasible
for a small number of objects, it is not easy to apply RM to large AGN samples
because RM is fairly time-consuming (always continuing for months to years).
Some multi-object RM campaigns based on fiber-fed telescopes, e.g., the RM
campaigns of SDSS \citep{shen2015} and OzDES \citep{king2015}, are committed to
enlarge the sample of RM objects but still ongoing. Fortunately, the \rl\
relationship can be used to obtain \rblr\ from the single-epoch spectra very
simply. 

The geometry and kinematics of the BLRs determine the virial factor \fblr\ in BH
mass estimate (in Equation \ref{eqn:bh_mass}). Comparing the RM AGNs with
stellar velocity dispersion measurements with $\mbh-\sigma_{*}$ relation of
inactive galaxies gives $\fblr \sim 1$ if the velocity width of \hb\ is measured
from \fwhmhb\ \citep[e.g.,][]{onken2004, ho2014, woo2015}. The virial factor for
high-\Rfe/high-accretion-rate AGNs is still a matter of some debate. 
Through fitting AGN spectral energy distribution by accretion disk model, 
\cite{mejia-restrepo2018} found a correlation between the virial factor and 
the width of emission line ($\fblr\sim3$ if $\fwhmhb\sim1500\,{\rm km\ s^{-1}}$ and 
$\fblr\sim0.4$ if $\fwhmhb\sim10000\,{\rm km\ s^{-1}}$).
\cite{yu2019} also shows potential a correlation between \fblr\ and line width. 
From the
BLR modeling results of the small samples in \cite{pancoast2014, grier2017,
williams2018}, the virial factor is roughly consistent with the value derived
from the $\mbh-\sigma_{*}$ relation, and the virial factors of individual objects
do no show significant correlation with the Eddington ratios 
(or show potential and weak
anti-correlation, namely smaller virial factor for higher
Eddington ratio). NLS1s (thought to have smaller BH masses and higher accretion
rates) tend to host pseudobulges \citep[e.g.,][]{mathur2012}. \cite{ho2014}
classified the AGN sample to classical bulges/ellipticals and pseudobulges, and
derived a virial factor of the AGNs with pseudobulges smaller than 1.
\cite{woo2015} found that NLS1s have no significant differences from the other
AGNs, and derived $\fblr=1.12$ if using \fwhmhb\ as the line-width measurement.
Thus, adopting $\fblr\sim1$ and acknowledging its large uncertainty is
acceptable at present. 

As a simple test, we compare the BH masses measured by RM ($M_{\bullet, {\rm
RM}}$) with the masses estimated from the single-epoch spectra ($M_{\bullet,
{\rm SE}}$) using both of the new scaling relation in Section
\ref{sec:new_scaling_relation} and the traditional \rl\ relationship (see
Section \ref{sec:deviation}) in Figure \ref{fig:mass_mass}. Here, we adopt
$\fblr=1$ for simplicity and list the RM BH masses in Table
\ref{tab:rm_results}. The scatter is quantified by the standard deviation of
$M_{\bullet, {\rm SE}} / M_{\bullet, {\rm RM}}$. We do not plot the error bars
in order to show clearly the differences between the estimates from the new
scaling relation and the \rl\ relationship. The scatter of $M_{\bullet, {\rm
SE}}$ from the new scaling relation is 0.52, while the scatter obtained by the
traditional
\rl\ relationship is 1.18. The embedded plot in Figure \ref{fig:mass_mass} 
shows the distributions of $M_{\bullet, {\rm SE}} / M_{\bullet, {\rm RM}}$. The
$M_{\bullet, {\rm SE}} / M_{\bullet, {\rm RM}}$ distribution is biased toward
the value larger than 1 for the traditional \rl\ relationship, which means the
BH masses are overestimated if using this simple relationship. 

Recently, \cite{MaryLoli2019} found a correction for the time delay based on the
dimensionless accretion rate (\dotm\ in the following Section
\ref{sec:accretion_rate}) considering the anti-correlation between \fblr\ and
line width, established a correlation between the corrected time lag ($R_{\rm
BLR}^{\rm corr}$) and $L_{5100}$, and discussed the measurements of the
cosmological distances using this correlation. Their correction ($R_{\rm
BLR}^{\rm corr}$) relies on the measured $L_{5100}$, \fwhmhb, and \rblr\ itself.
The new scaling relation in the present paper is established in a 
different perspective (based on spectral properties), and can 
deduce \rblr\ directly from $L_{5100}$ and \Rfe.

\subsection{Accretion Rate and Eddington Ratio}
\label{sec:accretion_rate}

Because of the shortened lags, the accretion rates or Eddington ratios would be
underestimated by factors of a few if using the traditional \rl\ relationship.
From the standard disk model \citep{shakura1973}, the
accretion rate can be estimated by the formula of 
\begin{equation}
    \label{eqn:mdot}
    \dotm = 20.1 \left(\frac{\ell_{44}}{\cos i}\right)^{3/2} m^{-2}_7,
\end{equation}
where $m_7 = \mbh / 10^7 M_{\odot}$, and $i$ is inclination angle of the
accretion disk (here we adopt $\cos i = 0.75$ as an average value for all of the
AGNs, see the discussions in \citealt{du2014, wang2014, du2016V}). Here, we
compare the \dotm\ and Eddington ratio
($L_{\rm Bol}/L_{\rm Edd}$) estimates obtained from the single-epoch spectra
(using the new scaling relation and the \rl\ relationship) with those values
calculated from RM (listed in Table \ref{tab:rm_results}) in Figure \ref{fig:mdot_mdot}, where $L_{\rm Bol}$ is the
bolometric luminosity and $L_{\rm Edd}=1.5\times10^{38}(\mbh/M_{\odot})$ is the
Eddington luminosity for the gas with solar composition. Here we simply adopt
$L_{\rm Bol}=10\,L_{5100}$ \citep{kaspi2000}, but it should be noted that the
bolometric correction factor depends on accretion rate or BH mass
\citep{jin2012}. We do not draw the error bars in order to show the differences
at high-\dotm\ end more clearly. It is obvious that the points of the new
scaling relation at high-accretion-rate end are much closer to the diagonal,
while those estimated by the traditional \rl\ relationship are biased downward.
The distributions and the standard deviations of $\log\,(\dotm_{\rm
SE}/\dotm_{\rm RM})$ and $\log\,[(L_{\rm Bol}/L_{\rm Edd})_{\rm SE}/(L_{\rm
Bol}/L_{\rm Edd})_{\rm RM}]$ are also provided in Figure \ref{fig:mdot_mdot}.
The new scaling relation should be preferentially used in the estimation of
accretion rates or Eddington ratios from the single-epoch spectra in the
statistical study of large AGN samples.

\cite{du2016FP} established a bivariate correlation between \dotm\ and (\Rfe,
\Dhb), which can be used to estimate the accretion rate directly from the BLR
properties. It is called the fundamental plane (FP) of BLR. The FP can deduce
\dotm\ or Eddington ratio estimates without any luminosity measurements (see also 
\citealt{negrete2018}), however
has fairly large uncertainties (the scatter of the FP is 0.7 for \dotm\ and 0.48
for Eddington ratio, respectively, see more details in \citealt{du2016FP}). As a
comparison, we plot the \dotm\ estimates from the FP and the new scaling
relation in Figure \ref{fig:mdot_mdot_fp}. The single-epoch \dotm\ and
Eddington ratios corresponding to the FP method are estimated from the \Rfe\ and
\Dhb\ listed in Table \ref{tab:properties} and the FP in \cite{du2016FP}. It
should be noted that the we switch x and y axes of Figure \ref{fig:mdot_mdot_fp}
(with respect to Figure \ref{fig:mdot_mdot}) for easier comparison with the
figures in \cite{du2016FP}. Again, we do not draw the error bars of the \dotm\
from the new scaling relation for clarity. The scatters of the \dotm\ and
$L_{\rm Bol}/L_{\rm Edd}$ estimated from the FP are much larger. The FP connect
the BLR physics with the accretion status of AGNs, however the different
temperature, number density, metallicity of the BLR in different AGN introduces
large uncertainties. The FP \citep{du2016FP} is a good beginning that searches
for direct indicator of accretion rate from the BLR observational properties,
but its scatter and accuracy should be improved by including more observational
properties in the future. 

The strong correlation between accretion rates \dotm\ and \Rfe\ has been
explored by \cite{panda2018, panda2019}, but still remains open. Accretion flows to the
central BHs supplied by either star formation from torus \citep{wang2010}, or
asymmetric dynamics \citep{begelman2009} will have different dependence on
metallicity, however, \Rfe\ is not a unique function of metallicity
\citep{baldwin2004, verner2004}. More details of photoionization are necessary
to investigate \Rfe\ dependence on BLR clouds (density, temperature and
metallicity) and the SEDs of accretion disks.  

\subsection{Accretion Rate or Orientation?}
\label{sec:orientation}

\begin{figure}
    \centering
    \includegraphics[width=0.47\textwidth]{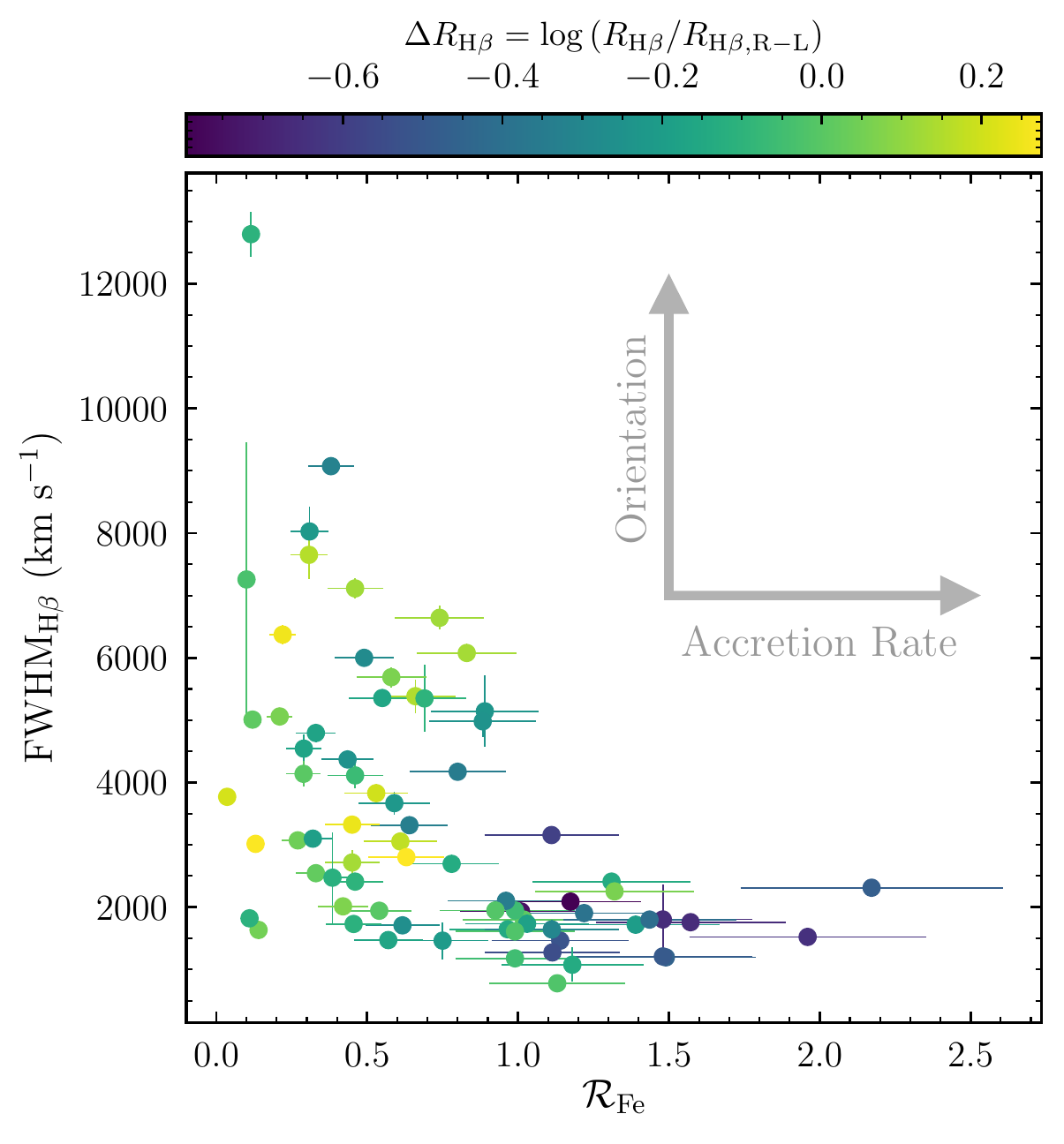}
    \caption{Main Sequence of AGNs. The points are color-coded by $\Delta \rhb$.
    Accretion rate/Eddington ratio drives the variation of \Rfe, while the
    orientation effect dominantly controls the dispersion in \fwhmhb\ at fixed
    \Rfe\ \citep{marziani2001, shen2014}. It is obvious that accretion rate definitely plays
    the primary role in the shortening of the time lags (see more details in
    Section \ref{sec:orientation}). }
    \label{fig:eigenvector1}
\end{figure}

The eigenvector 1 sequence can do some help
to break the degeneracy of accretion rate and orientation.
It was demonstrated that the
accretion rate or Eddington ratio drives the variation of \Rfe, while the
orientation effect dominantly controls the dispersion in \fwhmhb\ at fixed \Rfe\ 
\citep[e.g.,][]{marziani2001, shen2014}. 
We plot the
eigenvector 1 sequence of the RM sample color-coded by $\Delta \rhb$ in Figure
\ref{fig:eigenvector1}. It is obvious that the objects with the most significant
lag deviations are located in the lower right corner (with the strongest \Rfe).
Furthermore, there is no significant trend in the \fwhmhb-axis at fixed \Rfe. It
means that the accretion rate definitely plays the primary role in the
shortening of the time lags but the orientation does not contribute much. In
Figure \ref{fig:deltaR}, the dispersion of the $\Delta \rhb$ at lower \fwhmhb\
is larger, this is caused by the higher \Rfe\ in those objects, which is clearly
shown in Figure \ref{fig:eigenvector1}.

\section{Summary}
\label{sec:summary}

In this paper, we systematically investigate the dependence of the \rl\
relationship on optical spectra for a wide range of AGN parameters. The
reverberation mapping campaign of the AGNs with high accretion rates show many
objects deviate significantly from the traditional \rl\ relationship
\citep{du2015, du2016V, du2018}. We collect 8 different single-epoch spectral
properties to investigate how the deviation of an AGN from the \rl\
relationship correlates with those properties and to understand what is the
origin of the shortened lags.

\begin{itemize}

    \item The flux ratio between \feii\ and \hb\ lines (\Rfe) is confirmed to be
    the most prominent property that correlates with the deviation of an AGN
    from the \rl\ relationship. \Rfe\ is thought to be the indicator of
    accretion rate, therefore, accretion rate is the driver for the shortened
    lags. \fwhmhb, which is induced by the orientation of an AGN to line of
    sight, does not show clear trend with the lag shortening. Thus, the
    orientation is not a dominant factor.

    \item We established a new scaling relation with the form of $\log \rhb =
    \alpha + \beta \log \ell_{44} + \gamma \Rfe$, which can be used as a
    single-epoch estimator of BH mass and accretion rate, where 
    $\alpha = 1.65\pm0.06$, $\beta=0.45\pm0.03$, and $\gamma = -0.35\pm0.08$.
    The scatter of the new scaling relation is 0.196.

\end{itemize}

The new scaling relation provides an empirical relation of the BLR regions with
optical spectra. It is less biased and provides more robust estimates of the BH
mass and accretion rate/Eddington ratio.

\acknowledgements
JMW is grateful to all members of the SEAMBH project for great efforts since
2012. In particular, the continuous supports from the Lijiang 2.4m telescope of
Yunnan Observatories is acknowledged as well as staff's operation. Funding for
the telescope has been provided by CAS and the People's Government of Yunnan
Province. We acknowledge the support by National Key R\&D Program of China
(2016YFA0400701), by NSFC through grants
NSFC-11873048, −11833008, 11690024, and by grant No. QYZDJ-SSW-SLH007 from the
Key Research Program of Frontier Sciences, CAS, by the Strategic Priority
Research Program of the Chinese Academy of Sciences grant No. XDB23010400.

\startlongtable
\begin{deluxetable*}{lcccccccccc}
\tablecolumns{11}
\setlength{\tabcolsep}{1pt}
\tablecaption{Reverberation Mapped AGNs and Their Results\label{tab:rm_results}}
\tabletypesize{\scriptsize}
\tablehead{
    \colhead{Objects}                             &
    \colhead{$\tau_{\hb}$}                        &
    \colhead{$\log L_{5100}$}                     &
    \colhead{FWHM}                                &
    \colhead{$\log\left(\mbh/{M_{\odot}}\right)$} &
    \colhead{$\log\dotm$}                         &
    \colhead{$\log L_{\hb}$}                      &
    \colhead{EW(\hb)}                             &
    \colhead{$\log L_{\oiii}$}                    &
    \colhead{EW(\oiii)}                           &
    \colhead{Ref.}                                \\ \cline{2-11}
    \colhead{}                                    &
    \colhead{(days)}                              &
    \colhead{(erg s$^{-1}$)}                      &
    \colhead{(km s$^{-1}$)}                       &
    \colhead{}                                    &
    \colhead{}                                    &
    \colhead{(erg s$^{-1}$)}                      &
    \colhead{(\AA)}                               &
    \colhead{(erg s$^{-1}$)}                      &
    \colhead{(\AA)}                               &
    \colhead{}
}
\startdata
         Mrk 335 &     $  8.7_{-  1.9}^{+  1.6}$  &     $43.69\pm 0.06$  &     $ 2096\pm  170$  &     $6.87_{-0.14}^{+0.10}$  &     $ 1.17_{- 0.30}^{+ 0.31}$  &     $42.03\pm 0.06$  &     $110.5\pm 22.3$  &     $41.49\pm 0.06$  &     $ 31.8\pm  2.4$ & 1, 2, 3      \\
                 &     $ 16.8_{-  4.2}^{+  4.8}$  &     $43.76\pm 0.06$  &     $ 1792\pm    3$  &     $7.02_{-0.12}^{+0.11}$  &     $ 1.28_{- 0.29}^{+ 0.30}$  &     $42.13\pm 0.06$  &     $119.7\pm 23.3$  &     $41.62\pm 0.06$  &     $ 36.9\pm  1.9$ & 4, 5, 6, 7   \\
                 &     $ 12.5_{-  5.5}^{+  6.6}$  &     $43.84\pm 0.06$  &     $ 1679\pm    2$  &     $6.84_{-0.25}^{+0.18}$  &     $ 1.39_{- 0.29}^{+ 0.30}$  &     $42.18\pm 0.06$  &     $111.2\pm 21.1$  &     $41.62\pm 0.06$  &     $ 31.1\pm  1.6$ & 4, 5, 6, 7   \\
                 &     $ 14.3_{-  0.7}^{+  0.7}$  &     $43.74\pm 0.06$  &     $ 1724\pm  236$  &     $6.92_{-0.14}^{+0.11}$  &     $ 1.25_{- 0.29}^{+ 0.30}$  &     $41.99\pm 0.07$  &     $ 89.5\pm 19.5$  &     $41.62\pm 0.06$  &     $ 38.6\pm  2.5$ & 4, 8, 9      \\
                 & $\bf{ 14.0_{-  3.4}^{+  4.6}}$ & $\bf{43.76\pm 0.07}$ & $\bf{ 1707\pm   79}$ & $\bf{6.93_{-0.11}^{+0.10}}$ & $\bf{ 1.27_{- 0.17}^{+ 0.18}}$ & $\bf{42.09\pm 0.09}$ & $\bf{108.2\pm 16.8}$ & $\bf{41.55\pm 0.10}$ & $\bf{ 34.5\pm  3.8}$& ...          \\
     PG 0026+129 &     $111.0_{- 28.3}^{+ 24.1}$  &     $44.97\pm 0.02$  &     $ 2544\pm   56$  &     $8.15_{-0.13}^{+0.09}$  &     $ 0.65_{- 0.20}^{+ 0.28}$  &     $42.93\pm 0.04$  &     $ 46.2\pm  4.7$  &     $42.60\pm 0.02$  &     $ 21.6\pm  1.2$ & 4, 5, 6, 10  \\
     PG 0052+251 &     $ 89.8_{- 24.1}^{+ 24.5}$  &     $44.81\pm 0.03$  &     $ 5008\pm   73$  &     $8.64_{-0.14}^{+0.11}$  &     $-0.59_{- 0.25}^{+ 0.31}$  &     $43.13\pm 0.05$  &     $107.4\pm 14.8$  &     $42.79\pm 0.02$  &     $ 48.9\pm  3.6$ & 4, 5, 6, 10  \\
        Fairall9 &     $ 17.4_{-  4.3}^{+  3.2}$  &     $43.98\pm 0.04$  &     $ 5999\pm   66$  &     $8.09_{-0.12}^{+0.07}$  &     $-0.71_{- 0.21}^{+ 0.31}$  &     $42.67\pm 0.04$  &     $249.8\pm 32.0$  &     $42.15\pm 0.03$  &     $ 75.8\pm  4.5$ & 4, 5, 6, 11  \\
         Mrk 590 &     $ 20.7_{-  2.7}^{+  3.5}$  &     $43.59\pm 0.06$  &     $ 2788\pm   29$  &     $7.50_{-0.06}^{+0.07}$  &     $-0.22_{- 0.25}^{+ 0.24}$  &     $41.92\pm 0.06$  &     $107.0\pm 22.0$  &     $41.30\pm 0.06$  &     $ 25.8\pm  2.1$ & 4, 5, 6, 7   \\
                 &     $ 14.0_{-  8.8}^{+  8.5}$  &     $43.14\pm 0.09$  &     $ 3729\pm  426$  &     $7.58_{-0.48}^{+0.22}$  &     $-0.91_{- 0.30}^{+ 0.28}$  &     $41.58\pm 0.16$  &     $142.8\pm 62.2$  &     $41.30\pm 0.06$  &     $ 74.4\pm 13.2$ & 4, 5, 6, 7   \\
                 &     $ 29.2_{-  5.0}^{+  4.9}$  &     $43.38\pm 0.07$  &     $ 2744\pm   79$  &     $7.63_{-0.09}^{+0.07}$  &     $-0.54_{- 0.26}^{+ 0.25}$  &     $41.75\pm 0.07$  &     $119.7\pm 28.1$  &     $41.30\pm 0.06$  &     $ 42.3\pm  4.7$ & 4, 5, 6, 7   \\
                 &     $ 28.8_{-  4.2}^{+  3.6}$  &     $43.65\pm 0.06$  &     $ 2500\pm   43$  &     $7.55_{-0.07}^{+0.05}$  &     $-0.13_{- 0.25}^{+ 0.24}$  &     $41.92\pm 0.07$  &     $ 94.9\pm 20.7$  &     $41.30\pm 0.06$  &     $ 22.8\pm  1.8$ & 4, 5, 6, 7   \\
                 & $\bf{ 25.6_{-  5.3}^{+  6.5}}$ & $\bf{43.50\pm 0.21}$ & $\bf{ 2716\pm  202}$ & $\bf{7.55_{-0.08}^{+0.07}}$ & $\bf{-0.41_{- 0.36}^{+ 0.36}}$ & $\bf{41.85\pm 0.12}$ & $\bf{108.6\pm 20.2}$ & $\bf{41.30\pm 0.06}$ & $\bf{ 29.3\pm 12.4}$& ...          \\
        Mrk 1044 &     $ 10.5_{-  2.7}^{+  3.3}$  &     $43.10\pm 0.10$  &     $ 1178\pm   22$  &     $6.45_{-0.13}^{+0.12}$  &     $ 1.22_{- 0.41}^{+ 0.40}$  &     $41.39\pm 0.09$  &     $101.4\pm 31.9$  &     $40.45\pm 0.09$  &     $ 11.6\pm  1.4$ & 1, 2, 3      \\
          3C 120 &     $ 38.1_{- 15.3}^{+ 21.3}$  &     $44.07\pm 0.05$  &     $ 2327\pm   48$  &     $7.61_{-0.22}^{+0.19}$  &     $ 0.03_{- 0.37}^{+ 0.37}$  &     $42.37\pm 0.06$  &     $100.9\pm 18.3$  &     $42.27\pm 0.05$  &     $ 81.2\pm  5.0$ & 4, 5, 6, 7   \\
                 &     $ 25.9_{-  2.3}^{+  2.3}$  &     $43.94\pm 0.05$  &     $ 3529\pm  176$  &     $7.80_{-0.06}^{+0.05}$  &     $-0.17_{- 0.37}^{+ 0.37}$  &     $42.36\pm 0.05$  &     $135.9\pm 22.3$  &     $42.36\pm 0.05$  &     $134.0\pm  5.1$ & 4, 8, 9      \\
                 & $\bf{ 26.2_{-  6.6}^{+  8.7}}$ & $\bf{44.00\pm 0.10}$ & $\bf{ 2472\pm  729}$ & $\bf{7.79_{-0.15}^{+0.15}}$ & $\bf{-0.07_{- 0.30}^{+ 0.30}}$ & $\bf{42.36\pm 0.04}$ & $\bf{118.8\pm 28.9}$ & $\bf{42.32\pm 0.07}$ & $\bf{116.3\pm 41.4}$& ...          \\
 IRAS 04416+1215 &     $ 13.3_{-  1.4}^{+ 13.9}$  &     $44.47\pm 0.03$  &     $ 1522\pm   44$  &     $6.78_{-0.06}^{+0.31}$  &     $ 2.63_{- 0.67}^{+ 0.16}$  &     $42.51\pm 0.02$  &     $ 55.8\pm  4.7$  &     $42.13\pm 0.05$  &     $ 23.6\pm  2.6$ & 1, 2, 3      \\
         Ark 120 &     $ 47.1_{- 12.4}^{+  8.3}$  &     $43.98\pm 0.06$  &     $ 6042\pm   35$  &     $8.53_{-0.13}^{+0.07}$  &     $-1.48_{- 0.23}^{+ 0.24}$  &     $42.60\pm 0.05$  &     $211.5\pm 37.5$  &     $41.54\pm 0.05$  &     $ 18.5\pm  1.6$ & 4, 5, 6, 7   \\
                 &     $ 37.1_{-  5.4}^{+  4.8}$  &     $43.63\pm 0.08$  &     $ 6246\pm   78$  &     $8.45_{-0.07}^{+0.05}$  &     $-2.01_{- 0.27}^{+ 0.27}$  &     $42.43\pm 0.07$  &     $321.1\pm 77.6$  &     $41.54\pm 0.05$  &     $ 41.5\pm  6.5$ & 4, 5, 6, 7   \\
                 & $\bf{ 39.5_{-  7.8}^{+  8.5}}$ & $\bf{43.87\pm 0.25}$ & $\bf{ 6077\pm  147}$ & $\bf{8.47_{-0.08}^{+0.07}}$ & $\bf{-1.70_{- 0.41}^{+ 0.41}}$ & $\bf{42.54\pm 0.13}$ & $\bf{244.8\pm 80.3}$ & $\bf{41.54\pm 0.05}$ & $\bf{ 22.3\pm 12.9}$& ...          \\
  MCG +08-11-011 &     $ 15.7_{-  0.5}^{+  0.5}$  &     $43.33\pm 0.11$  &     $ 4139\pm  207$  &     $7.72_{-0.05}^{+0.04}$  &     $-0.96_{- 0.28}^{+ 0.25}$  &     $41.66\pm 0.09$  &     $108.7\pm 35.3$  &     $42.06\pm 0.07$  &     $276.8\pm 54.3$ & 12           \\
         Mrk 374 &     $ 14.8_{-  3.3}^{+  5.8}$  &     $43.77\pm 0.04$  &     $ 4980\pm  249$  &     $7.86_{-0.12}^{+0.15}$  &     $-0.56_{- 0.36}^{+ 0.30}$  &     $41.83\pm 0.04$  &     $ 58.3\pm  7.9$  &     $41.59\pm 0.04$  &     $ 33.7\pm  1.8$ & 12           \\
          Mrk 79 &     $  9.0_{-  7.8}^{+  8.3}$  &     $43.63\pm 0.07$  &     $ 5056\pm   85$  &     $7.65_{-0.88}^{+0.28}$  &     $-0.75_{- 0.34}^{+ 0.41}$  &     $41.89\pm 0.07$  &     $ 92.4\pm 21.0$  &     $41.67\pm 0.07$  &     $ 56.1\pm  3.8$ & 4, 5, 6, 7   \\
                 &     $ 16.1_{-  6.6}^{+  6.6}$  &     $43.74\pm 0.07$  &     $ 4760\pm   31$  &     $7.85_{-0.23}^{+0.15}$  &     $-0.59_{- 0.34}^{+ 0.41}$  &     $41.92\pm 0.08$  &     $ 78.1\pm 18.3$  &     $41.67\pm 0.07$  &     $ 44.0\pm  2.9$ & 4, 5, 6, 7   \\
                 &     $ 16.0_{-  5.8}^{+  6.4}$  &     $43.66\pm 0.07$  &     $ 4766\pm   71$  &     $7.85_{-0.20}^{+0.15}$  &     $-0.70_{- 0.34}^{+ 0.41}$  &     $41.89\pm 0.07$  &     $ 86.0\pm 19.3$  &     $41.67\pm 0.07$  &     $ 51.9\pm  3.5$ & 4, 5, 6, 7   \\
                 & $\bf{ 15.6_{-  4.9}^{+  5.1}}$ & $\bf{43.68\pm 0.07}$ & $\bf{ 4793\pm  145}$ & $\bf{7.84_{-0.16}^{+0.12}}$ & $\bf{-0.68_{- 0.21}^{+ 0.25}}$ & $\bf{41.90\pm 0.05}$ & $\bf{ 85.4\pm 13.3}$ & $\bf{41.67\pm 0.07}$ & $\bf{ 50.2\pm  6.6}$& ...          \\
    SDSS J074352 &     $ 43.9_{-  4.2}^{+  5.2}$  &     $45.37\pm 0.02$  &     $ 3156\pm   36$  &     $7.93_{-0.04}^{+0.05}$  &     $ 1.69_{- 0.13}^{+ 0.12}$  &     $43.48\pm 0.01$  &     $ 65.8\pm  3.5$  &     $42.42\pm 0.05$  &     $  5.6\pm  0.7$ & 13           \\
    SDSS J075051 &     $ 66.6_{-  9.9}^{+ 18.7}$  &     $45.33\pm 0.01$  &     $ 1904\pm    9$  &     $7.67_{-0.07}^{+0.11}$  &     $ 2.14_{- 0.24}^{+ 0.16}$  &     $43.34\pm 0.03$  &     $ 51.9\pm  4.5$  &     $42.37\pm 0.02$  &     $  5.6\pm  0.3$ & 13           \\
    SDSS J075101 &     $ 33.4_{-  5.6}^{+ 15.6}$  &     $44.12\pm 0.05$  &     $ 1495\pm   67$  &     $7.16_{-0.09}^{+0.17}$  &     $ 1.30_{- 0.24}^{+ 0.24}$  &     $42.25\pm 0.03$  &     $ 68.1\pm  8.6$  &     $41.81\pm 0.03$  &     $ 25.0\pm  3.1$ & 9            \\
                 &     $ 28.6_{-  6.8}^{+  5.6}$  &     $44.24\pm 0.04$  &     $ 1679\pm   35$  &     $7.20_{-0.12}^{+0.08}$  &     $ 1.47_{- 0.23}^{+ 0.23}$  &     $42.38\pm 0.04$  &     $ 70.4\pm  9.0$  &     $41.75\pm 0.03$  &     $ 16.5\pm  1.8$ & 13           \\
                 & $\bf{ 30.4_{-  5.8}^{+  7.3}}$ & $\bf{44.18\pm 0.09}$ & $\bf{ 1645\pm  139}$ & $\bf{7.18_{-0.09}^{+0.08}}$ & $\bf{ 1.39_{- 0.21}^{+ 0.21}}$ & $\bf{42.30\pm 0.10}$ & $\bf{ 69.2\pm  6.4}$ & $\bf{41.77\pm 0.05}$ & $\bf{ 19.8\pm  6.0}$& ...          \\
         Mrk 382 &     $  7.5_{-  2.0}^{+  2.9}$  &     $43.12\pm 0.08$  &     $ 1462\pm  296$  &     $6.50_{-0.29}^{+0.19}$  &     $ 1.18_{- 0.53}^{+ 0.69}$  &     $41.01\pm 0.05$  &     $ 39.6\pm  9.0$  &     $40.93\pm 0.05$  &     $ 32.5\pm  5.9$ & 1, 2, 3      \\
    SDSS J075949 &     $ 55.0_{- 13.1}^{+ 17.0}$  &     $44.20\pm 0.03$  &     $ 1807\pm   11$  &     $7.54_{-0.12}^{+0.12}$  &     $ 0.90_{- 0.56}^{+ 0.56}$  &     $42.48\pm 0.02$  &     $ 97.5\pm  9.1$  &     $41.44\pm 0.04$  &     $  8.9\pm  1.1$ & 14           \\
                 &     $ 26.4_{-  9.5}^{+ 11.6}$  &     $44.19\pm 0.06$  &     $ 1783\pm   17$  &     $7.21_{-0.19}^{+0.16}$  &     $ 0.88_{- 0.61}^{+ 0.60}$  &     $42.47\pm 0.04$  &     $ 98.9\pm 17.0$  &     $41.31\pm 0.07$  &     $  6.8\pm  1.5$ & 13           \\
                 & $\bf{ 43.9_{- 19.0}^{+ 33.1}}$ & $\bf{44.20\pm 0.03}$ & $\bf{ 1800\pm   19}$ & $\bf{7.44_{-0.25}^{+0.25}}$ & $\bf{ 0.89_{- 0.41}^{+ 0.41}}$ & $\bf{42.48\pm 0.02}$ & $\bf{ 97.8\pm  8.1}$ & $\bf{41.40\pm 0.10}$ & $\bf{  8.3\pm  1.9}$& ...          \\
    SDSS J080101 &     $  8.3_{-  2.7}^{+  9.7}$  &     $44.27\pm 0.03$  &     $ 1930\pm   18$  &     $6.78_{-0.17}^{+0.34}$  &     $ 2.33_{- 0.72}^{+ 0.39}$  &     $42.58\pm 0.02$  &     $105.5\pm  8.3$  &     $41.16\pm 0.05$  &     $  3.9\pm  0.5$ & 9            \\
    SDSS J080131 &     $ 11.5_{-  3.6}^{+  8.4}$  &     $43.98\pm 0.04$  &     $ 1188\pm    3$  &     $6.50_{-0.16}^{+0.24}$  &     $ 2.45_{- 0.51}^{+ 0.40}$  &     $42.08\pm 0.03$  &     $ 64.0\pm  7.0$  &     $41.41\pm 0.07$  &     $ 13.6\pm  2.5$ & 9            \\
                 &     $ 11.2_{-  9.8}^{+ 14.8}$  &     $43.95\pm 0.04$  &     $ 1290\pm   13$  &     $6.56_{-0.90}^{+0.37}$  &     $ 2.40_{- 0.51}^{+ 0.41}$  &     $41.96\pm 0.05$  &     $ 52.3\pm  7.7$  &     $41.56\pm 0.04$  &     $ 20.9\pm  2.9$ & 14           \\
                 & $\bf{ 11.5_{-  3.7}^{+  7.5}}$ & $\bf{43.97\pm 0.04}$ & $\bf{ 1194\pm   70}$ & $\bf{6.51_{-0.17}^{+0.22}}$ & $\bf{ 2.43_{- 0.36}^{+ 0.29}}$ & $\bf{42.05\pm 0.09}$ & $\bf{ 59.5\pm 10.0}$ & $\bf{41.52\pm 0.12}$ & $\bf{ 17.9\pm  5.8}$& ...          \\
     PG 0804+761 &     $146.9_{- 18.9}^{+ 18.8}$  &     $44.91\pm 0.02$  &     $ 3053\pm   38$  &     $8.43_{-0.06}^{+0.05}$  &     $ 0.00_{- 0.13}^{+ 0.15}$  &     $43.29\pm 0.03$  &     $122.5\pm 10.3$  &     $42.42\pm 0.03$  &     $ 16.6\pm  0.9$ & 4, 5, 6, 10  \\
    SDSS J081441 &     $ 18.4_{-  8.4}^{+ 12.7}$  &     $44.01\pm 0.07$  &     $ 1615\pm   22$  &     $6.97_{-0.27}^{+0.23}$  &     $ 1.14_{- 0.52}^{+ 0.51}$  &     $42.42\pm 0.03$  &     $132.0\pm 23.7$  &     $41.36\pm 0.07$  &     $ 11.4\pm  2.7$ & 9            \\
                 &     $ 26.8_{-  5.9}^{+  7.3}$  &     $43.95\pm 0.04$  &     $ 1782\pm   16$  &     $7.22_{-0.11}^{+0.10}$  &     $ 1.05_{- 0.47}^{+ 0.47}$  &     $42.39\pm 0.02$  &     $140.4\pm 16.2$  &     $41.29\pm 0.04$  &     $ 11.2\pm  1.4$ & 13           \\
                 & $\bf{ 25.3_{-  7.5}^{+ 10.4}}$ & $\bf{43.96\pm 0.06}$ & $\bf{ 1730\pm  121}$ & $\bf{7.18_{-0.20}^{+0.20}}$ & $\bf{ 1.09_{- 0.35}^{+ 0.35}}$ & $\bf{42.40\pm 0.03}$ & $\bf{137.9\pm 14.7}$ & $\bf{41.30\pm 0.06}$ & $\bf{ 11.2\pm  1.3}$& ...          \\
    SDSS J081456 &     $ 24.3_{- 16.4}^{+  7.7}$  &     $43.99\pm 0.04$  &     $ 2409\pm   61$  &     $7.44_{-0.49}^{+0.12}$  &     $ 0.59_{- 0.30}^{+ 1.03}$  &     $42.15\pm 0.03$  &     $ 74.4\pm  7.6$  &     $41.66\pm 0.04$  &     $ 24.2\pm  2.7$ & 9            \\
        NGC 2617 &     $  4.3_{-  1.4}^{+  1.1}$  &     $42.67\pm 0.16$  &     $ 8026\pm  401$  &     $7.74_{-0.17}^{+0.11}$  &     $-1.98_{- 0.51}^{+ 0.55}$  &     $41.18\pm 0.12$  &     $165.8\pm 74.9$  &     $40.57\pm 0.10$  &     $ 41.0\pm 11.4$ & 12           \\
    SDSS J083553 &     $ 12.4_{-  5.4}^{+  5.4}$  &     $44.44\pm 0.02$  &     $ 1758\pm   16$  &     $6.87_{-0.25}^{+0.16}$  &     $ 2.41_{- 0.35}^{+ 0.53}$  &     $42.48\pm 0.02$  &     $ 56.1\pm  4.0$  &     $41.84\pm 0.03$  &     $ 12.6\pm  1.0$ & 13           \\
    SDSS J084533 &     $ 15.2_{-  6.3}^{+  3.2}$  &     $44.54\pm 0.04$  &     $ 1243\pm   13$  &     $6.66_{-0.23}^{+0.08}$  &     $ 2.77_{- 0.34}^{+ 0.35}$  &     $42.58\pm 0.05$  &     $ 55.9\pm  7.5$  &     $41.85\pm 0.04$  &     $ 10.4\pm  1.3$ & 14           \\
                 &     $ 19.9_{-  3.9}^{+  7.3}$  &     $44.52\pm 0.02$  &     $ 1297\pm   12$  &     $6.82_{-0.10}^{+0.14}$  &     $ 2.75_{- 0.32}^{+ 0.33}$  &     $42.60\pm 0.03$  &     $ 61.7\pm  5.1$  &     $41.75\pm 0.03$  &     $  8.6\pm  0.7$ & 13           \\
                 & $\bf{ 18.1_{-  4.7}^{+  6.0}}$ & $\bf{44.53\pm 0.02}$ & $\bf{ 1273\pm   39}$ & $\bf{6.76_{-0.15}^{+0.14}}$ & $\bf{ 2.76_{- 0.24}^{+ 0.24}}$ & $\bf{42.60\pm 0.03}$ & $\bf{ 60.0\pm  5.9}$ & $\bf{41.78\pm 0.07}$ & $\bf{  9.1\pm  1.4}$& ...          \\
     PG 0844+349 &     $ 32.3_{- 13.4}^{+ 13.7}$  &     $44.22\pm 0.07$  &     $ 2694\pm   58$  &     $7.66_{-0.23}^{+0.15}$  &     $ 0.50_{- 0.42}^{+ 0.57}$  &     $42.56\pm 0.05$  &     $111.2\pm 22.1$  &     $41.65\pm 0.03$  &     $ 13.9\pm  2.2$ & 5, 6, 10, 15 \\
    SDSS J085946 &     $ 34.8_{- 26.3}^{+ 19.2}$  &     $44.41\pm 0.03$  &     $ 1718\pm   16$  &     $7.30_{-0.61}^{+0.19}$  &     $ 1.51_{- 0.43}^{+ 1.27}$  &     $42.51\pm 0.02$  &     $ 63.1\pm  5.2$  &     $41.92\pm 0.04$  &     $ 16.4\pm  2.0$ & 14           \\
         Mrk 110 &     $ 24.3_{-  8.3}^{+  5.5}$  &     $43.68\pm 0.04$  &     $ 1543\pm    5$  &     $7.05_{-0.18}^{+0.09}$  &     $ 0.81_{- 0.32}^{+ 0.35}$  &     $42.12\pm 0.05$  &     $139.6\pm 20.4$  &     $41.87\pm 0.05$  &     $ 78.3\pm  5.4$ & 4, 5, 6, 7   \\
                 &     $ 20.4_{-  6.3}^{+ 10.5}$  &     $43.75\pm 0.04$  &     $ 1658\pm    3$  &     $7.04_{-0.16}^{+0.18}$  &     $ 0.92_{- 0.32}^{+ 0.34}$  &     $42.02\pm 0.05$  &     $ 94.8\pm 14.7$  &     $41.87\pm 0.05$  &     $ 66.2\pm  4.5$ & 4, 5, 6, 7   \\
                 &     $ 33.3_{- 10.0}^{+ 14.9}$  &     $43.53\pm 0.05$  &     $ 1600\pm   39$  &     $7.22_{-0.16}^{+0.16}$  &     $ 0.58_{- 0.33}^{+ 0.35}$  &     $41.97\pm 0.04$  &     $139.4\pm 20.5$  &     $41.87\pm 0.05$  &     $110.7\pm  8.3$ & 4, 5, 6, 7   \\
                 & $\bf{ 25.6_{-  7.2}^{+  8.9}}$ & $\bf{43.66\pm 0.12}$ & $\bf{ 1634\pm   83}$ & $\bf{7.10_{-0.14}^{+0.13}}$ & $\bf{ 0.77_{- 0.25}^{+ 0.26}}$ & $\bf{42.03\pm 0.08}$ & $\bf{123.8\pm 29.1}$ & $\bf{41.87\pm 0.05}$ & $\bf{ 81.5\pm 21.0}$& ...          \\
    SDSS J093302 &     $ 19.0_{-  4.3}^{+  3.8}$  &     $44.31\pm 0.13$  &     $ 1800\pm   25$  &     $7.08_{-0.11}^{+0.08}$  &     $ 1.79_{- 0.40}^{+ 0.40}$  &     $42.10\pm 0.05$  &     $ 31.8\pm 10.3$  &     $41.26\pm 0.06$  &     $  4.6\pm  1.5$ & 13           \\
    SDSS J093922 &     $ 11.9_{-  6.3}^{+  2.1}$  &     $44.07\pm 0.04$  &     $ 1209\pm   16$  &     $6.53_{-0.33}^{+0.07}$  &     $ 2.54_{- 0.20}^{+ 0.71}$  &     $42.09\pm 0.04$  &     $ 53.0\pm  6.7$  &     $41.31\pm 0.13$  &     $  8.8\pm  2.8$ & 9            \\
     PG 0953+414 &     $150.1_{- 22.6}^{+ 21.6}$  &     $45.19\pm 0.01$  &     $ 3071\pm   27$  &     $8.44_{-0.07}^{+0.06}$  &     $ 0.39_{- 0.14}^{+ 0.16}$  &     $43.29\pm 0.04$  &     $ 64.7\pm  5.9$  &     $42.73\pm 0.02$  &     $ 18.0\pm  1.0$ & 4, 5, 6, 10  \\
    SDSS J100402 &     $ 32.2_{-  4.2}^{+ 43.5}$  &     $45.52\pm 0.01$  &     $ 2088\pm    1$  &     $7.44_{-0.06}^{+0.37}$  &     $ 2.89_{- 0.75}^{+ 0.13}$  &     $43.54\pm 0.01$  &     $ 53.6\pm  1.3$  &     $42.45\pm 0.04$  &     $  4.4\pm  0.4$ & 13           \\
    SDSS J101000 &     $ 27.7_{-  7.6}^{+ 23.5}$  &     $44.76\pm 0.02$  &     $ 2311\pm    1$  &     $7.46_{-0.14}^{+0.27}$  &     $ 1.70_{- 0.56}^{+ 0.31}$  &     $42.77\pm 0.02$  &     $ 52.6\pm  3.4$  &     $41.50\pm 0.11$  &     $  2.8\pm  0.7$ & 13           \\
        NGC 3227 &     $  3.8_{-  0.8}^{+  0.8}$  &     $42.24\pm 0.11$  &     $ 4112\pm  206$  &     $7.09_{-0.12}^{+0.09}$  &     $-1.34_{- 0.36}^{+ 0.38}$  &     $40.38\pm 0.10$  &     $ 71.0\pm 23.6$  &     $40.68\pm 0.10$  &     $142.2\pm 22.1$ & 4, 9, 16     \\
    SDSS J102339 &     $ 24.9_{-  3.9}^{+ 19.8}$  &     $44.09\pm 0.03$  &     $ 1733\pm   29$  &     $7.16_{-0.08}^{+0.25}$  &     $ 1.29_{- 0.56}^{+ 0.20}$  &     $42.14\pm 0.03$  &     $ 57.0\pm  5.9$  &     $41.32\pm 0.04$  &     $  8.7\pm  1.0$ & 14           \\
         Mrk 142 &     $  7.9_{-  1.1}^{+  1.2}$  &     $43.56\pm 0.06$  &     $ 1588\pm   58$  &     $6.59_{-0.07}^{+0.07}$  &     $ 1.90_{- 0.86}^{+ 0.85}$  &     $41.60\pm 0.04$  &     $ 55.2\pm  9.5$  &     $40.86\pm 0.04$  &     $ 10.0\pm  1.3$ & 1, 2, 3      \\
                 &     $  2.7_{-  0.8}^{+  0.7}$  &     $43.61\pm 0.04$  &     $ 1462\pm    2$  &     $6.06_{-0.16}^{+0.10}$  &     $ 1.96_{- 0.82}^{+ 0.82}$  &     $41.66\pm 0.05$  &     $ 57.6\pm  8.6$  &     $41.24\pm 0.04$  &     $ 21.9\pm  1.4$ & 4, 17        \\
                 & $\bf{  6.4_{-  3.4}^{+  7.3}}$ & $\bf{43.59\pm 0.04}$ & $\bf{ 1462\pm   86}$ & $\bf{6.47_{-0.38}^{+0.38}}$ & $\bf{ 1.93_{- 0.59}^{+ 0.59}}$ & $\bf{41.62\pm 0.06}$ & $\bf{ 56.6\pm  6.6}$ & $\bf{41.06\pm 0.27}$ & $\bf{ 18.8\pm 10.5}$& ...          \\
        NGC 3516 &     $ 11.7_{-  1.5}^{+  1.0}$  &     $42.79\pm 0.20$  &     $ 5384\pm  269$  &     $7.82_{-0.08}^{+0.05}$  &     $-1.97_{- 0.52}^{+ 0.41}$  &     $41.06\pm 0.18$  &     $ 94.7\pm 59.2$  &     $40.85\pm 0.17$  &     $ 59.5\pm 18.2$ & 4, 9, 16     \\
   SBS 1116+583A &     $  2.3_{-  0.5}^{+  0.6}$  &     $42.14\pm 0.23$  &     $ 3668\pm  186$  &     $6.78_{-0.12}^{+0.11}$  &     $-0.87_{- 0.71}^{+ 0.51}$  &     $40.70\pm 0.07$  &     $186.8\pm104.1$  &     $40.50\pm 0.06$  &     $117.4\pm 61.1$ & 4, 17        \\
         Arp 151 &     $  4.0_{-  0.7}^{+  0.5}$  &     $42.55\pm 0.10$  &     $ 3098\pm   69$  &     $6.87_{-0.08}^{+0.05}$  &     $-0.44_{- 0.28}^{+ 0.30}$  &     $40.95\pm 0.11$  &     $130.0\pm 44.4$  &     $40.75\pm 0.07$  &     $ 80.7\pm 14.2$ & 4, 17        \\
        NGC 3783 &     $ 10.2_{-  2.3}^{+  3.3}$  &     $42.56\pm 0.18$  &     $ 3770\pm   68$  &     $7.45_{-0.11}^{+0.12}$  &     $-1.58_{- 0.59}^{+ 0.45}$  &     $41.01\pm 0.18$  &     $144.0\pm 83.7$  &     $40.95\pm 0.17$  &     $126.5\pm 14.6$ & 4, 5, 6, 18  \\
  MCG +06-26-012 &     $ 24.0_{-  4.8}^{+  8.4}$  &     $42.67\pm 0.11$  &     $ 1334\pm   80$  &     $6.92_{-0.12}^{+0.14}$  &     $-0.34_{- 0.45}^{+ 0.37}$  &     $41.03\pm 0.06$  &     $114.6\pm 32.5$  &     $40.55\pm 0.05$  &     $ 38.7\pm  8.6$ & 1, 2, 3      \\
       UGC 06728 &     $  1.4_{-  0.8}^{+  0.7}$  &     $41.86\pm 0.08$  &     $ 1642\pm  161$  &     $5.87_{-0.40}^{+0.19}$  &     $ 0.55_{- 0.51}^{+ 0.92}$  &     $39.85\pm 0.05$  &     $ 49.5\pm 10.8$  &     $39.69\pm 0.02$  &     $ 33.8\pm  6.6$ & 19           \\
        Mrk 1310 &     $  3.7_{-  0.6}^{+  0.6}$  &     $42.29\pm 0.14$  &     $ 2409\pm   24$  &     $6.62_{-0.08}^{+0.07}$  &     $-0.31_{- 0.39}^{+ 0.35}$  &     $40.56\pm 0.10$  &     $ 94.3\pm 38.2$  &     $41.05\pm 0.08$  &     $293.8\pm 84.8$ & 4, 17        \\
        NGC 4051 &     $  1.9_{-  0.5}^{+  0.5}$  &     $41.96\pm 0.19$  &     $  851\pm  277$  &     $5.42_{-0.53}^{+0.23}$  &     $ 0.99_{- 1.06}^{+ 1.11}$  &     $40.19\pm 0.18$  &     $ 86.8\pm 51.7$  &     $40.15\pm 0.17$  &     $ 79.2\pm 14.6$ & 4, 9, 16     \\
                 &     $  2.9_{-  1.3}^{+  0.9}$  &     $41.85\pm 0.18$  &     $ 1145\pm  192$  &     $5.87_{-0.37}^{+0.16}$  &     $ 0.82_{- 1.02}^{+ 1.09}$  &     $39.99\pm 0.18$  &     $ 71.7\pm 41.3$  &     $40.17\pm 0.17$  &     $108.7\pm  7.5$ & 12           \\
                 & $\bf{  2.1_{-  0.7}^{+  0.9}}$ & $\bf{41.90\pm 0.15}$ & $\bf{ 1076\pm  277}$ & $\bf{5.72_{-0.44}^{+0.34}}$ & $\bf{ 0.90_{- 0.74}^{+ 0.79}}$ & $\bf{40.09\pm 0.19}$ & $\bf{ 78.6\pm 34.3}$ & $\bf{40.16\pm 0.12}$ & $\bf{104.6\pm 24.4}$& ...          \\
        NGC 4151 &     $  6.6_{-  0.8}^{+  1.1}$  &     $42.09\pm 0.21$  &     $ 6371\pm  150$  &     $7.72_{-0.06}^{+0.07}$  &     $-2.81_{- 0.57}^{+ 0.37}$  &     $40.56\pm 0.20$  &     $150.8\pm100.6$  &     $41.60\pm 0.17$  &     $1637.3\pm439.0$& 4, 5, 6, 20  \\
     PG 1211+143 &     $ 93.8_{- 42.1}^{+ 25.6}$  &     $44.73\pm 0.08$  &     $ 2012\pm   37$  &     $7.87_{-0.26}^{+0.11}$  &     $ 0.84_{- 0.35}^{+ 0.63}$  &     $43.02\pm 0.06$  &     $100.2\pm 22.9$  &     $42.25\pm 0.03$  &     $ 16.9\pm  3.2$ & 5, 6, 10, 15 \\
         Mrk 202 &     $  3.0_{-  1.1}^{+  1.7}$  &     $42.26\pm 0.14$  &     $ 1471\pm   18$  &     $6.11_{-0.20}^{+0.20}$  &     $ 0.66_{- 0.65}^{+ 0.59}$  &     $40.40\pm 0.09$  &     $ 70.6\pm 27.5$  &     $40.50\pm 0.07$  &     $ 87.7\pm 25.8$ & 4, 17        \\
        NGC 4253 &     $  6.2_{-  1.2}^{+  1.6}$  &     $42.57\pm 0.12$  &     $ 1609\pm   39$  &     $6.49_{-0.10}^{+0.10}$  &     $ 0.36_{- 0.42}^{+ 0.36}$  &     $40.77\pm 0.12$  &     $ 81.1\pm 31.6$  &     $41.38\pm 0.12$  &     $326.8\pm 37.5$ & 4, 17        \\
     PG 1226+023 &     $146.8_{- 12.1}^{+  8.3}$  &     $45.92\pm 0.05$  &     $ 3314\pm   59$  &     $8.50_{-0.04}^{+0.03}$  &     $ 1.37_{- 0.14}^{+ 0.15}$  &     $44.11\pm 0.03$  &     $ 80.3\pm 10.3$  &     $43.24\pm 0.02$  &     $ 10.7\pm  1.3$ & 21           \\
     PG 1229+204 &     $ 37.8_{- 15.3}^{+ 27.6}$  &     $43.70\pm 0.05$  &     $ 3828\pm   54$  &     $8.03_{-0.23}^{+0.24}$  &     $-1.03_{- 0.55}^{+ 0.52}$  &     $42.31\pm 0.06$  &     $209.7\pm 38.3$  &     $41.75\pm 0.03$  &     $ 58.2\pm  6.1$ & 4, 5, 6, 10  \\
        NGC 4593 &     $  3.7_{-  0.8}^{+  0.8}$  &     $42.87\pm 0.18$  &     $ 5143\pm   16$  &     $7.28_{-0.10}^{+0.08}$  &     $-0.73_{- 0.52}^{+ 0.41}$  &     $41.17\pm 0.18$  &     $101.6\pm 59.0$  &     $40.64\pm 0.17$  &     $ 30.5\pm  3.2$ & 4, 6, 22     \\
                 &     $  4.3_{-  0.8}^{+  1.3}$  &     $42.38\pm 0.18$  &     $ 4395\pm  362$  &     $7.21_{-0.12}^{+0.13}$  &     $-1.47_{- 0.52}^{+ 0.41}$  &     $40.73\pm 0.18$  &     $115.4\pm 67.6$  &     $40.38\pm 0.18$  &     $ 51.8\pm  5.8$ & 9, 23        \\
                 & $\bf{  4.0_{-  0.7}^{+  0.8}}$ & $\bf{42.62\pm 0.37}$ & $\bf{ 5142\pm  572}$ & $\bf{7.26_{-0.09}^{+0.09}}$ & $\bf{-1.10_{- 0.64}^{+ 0.60}}$ & $\bf{40.95\pm 0.33}$ & $\bf{108.3\pm 45.7}$ & $\bf{40.51\pm 0.22}$ & $\bf{ 39.0\pm 14.9}$& ...          \\
IRAS F12397+3333 &     $  9.7_{-  1.8}^{+  5.5}$  &     $44.23\pm 0.05$  &     $ 1802\pm  560$  &     $6.79_{-0.45}^{+0.27}$  &     $ 2.26_{- 0.62}^{+ 0.98}$  &     $42.26\pm 0.04$  &     $ 54.2\pm  8.4$  &     $42.35\pm 0.04$  &     $ 66.9\pm  6.8$ & 1, 2, 3      \\
        NGC 4748 &     $  5.5_{-  2.2}^{+  1.6}$  &     $42.56\pm 0.12$  &     $ 1947\pm   66$  &     $6.61_{-0.23}^{+0.11}$  &     $ 0.10_{- 0.44}^{+ 0.61}$  &     $40.98\pm 0.10$  &     $136.8\pm 50.1$  &     $41.33\pm 0.10$  &     $300.3\pm 48.4$ & 4, 17        \\
     PG 1307+085 &     $105.6_{- 46.6}^{+ 36.0}$  &     $44.85\pm 0.02$  &     $ 5059\pm  133$  &     $8.72_{-0.26}^{+0.13}$  &     $-0.68_{- 0.28}^{+ 0.53}$  &     $43.13\pm 0.06$  &     $ 98.4\pm 15.1$  &     $42.73\pm 0.02$  &     $ 39.0\pm  2.2$ & 4, 5, 6, 10  \\
  MCG +06-30-015 &     $  5.3_{-  1.8}^{+  1.9}$  &     $41.65\pm 0.23$  &     $ 1958\pm   75$  &     $6.60_{-0.18}^{+0.13}$  &     $-1.28_{- 0.73}^{+ 0.58}$  &     $39.72\pm 0.13$  &     $ 60.0\pm 36.6$  &     $39.96\pm 0.12$  &     $104.9\pm 48.6$ & 24           \\
                 &     $  6.4_{-  2.7}^{+  3.1}$  &     $41.64\pm 0.12$  &     $ 1933\pm   81$  &     $6.67_{-0.24}^{+0.17}$  &     $-1.30_{- 0.44}^{+ 0.46}$  &     $39.97\pm 0.12$  &     $108.4\pm 42.8$  &     $39.98\pm 0.12$  &     $110.5\pm  7.8$ & 25           \\
                 & $\bf{  5.7_{-  1.7}^{+  1.8}}$ & $\bf{41.64\pm 0.11}$ & $\bf{ 1947\pm   58}$ & $\bf{6.63_{-0.15}^{+0.12}}$ & $\bf{-1.29_{- 0.38}^{+ 0.37}}$ & $\bf{39.85\pm 0.19}$ & $\bf{ 91.0\pm 48.6}$ & $\bf{39.97\pm 0.09}$ & $\bf{110.3\pm  8.7}$& ...          \\
        NGC 5273 &     $  2.2_{-  1.6}^{+  1.2}$  &     $41.54\pm 0.16$  &     $ 5688\pm  163$  &     $7.14_{-0.56}^{+0.19}$  &     $-2.50_{- 0.67}^{+ 1.33}$  &     $39.74\pm 0.11$  &     $ 82.2\pm 37.1$  &     $39.49\pm 0.08$  &     $ 46.5\pm 14.5$ & 26           \\
         Mrk 279 &     $ 16.7_{-  3.9}^{+  3.9}$  &     $43.71\pm 0.07$  &     $ 5354\pm   32$  &     $7.97_{-0.12}^{+0.09}$  &     $-0.89_{- 0.30}^{+ 0.33}$  &     $42.12\pm 0.06$  &     $132.2\pm 28.7$  &     $41.56\pm 0.06$  &     $ 36.9\pm  5.3$ & 4, 5, 6, 27  \\
     PG 1411+442 &     $124.3_{- 61.7}^{+ 61.0}$  &     $44.56\pm 0.02$  &     $ 2801\pm   43$  &     $8.28_{-0.30}^{+0.17}$  &     $-0.23_{- 0.38}^{+ 0.63}$  &     $42.85\pm 0.03$  &     $ 99.7\pm  8.2$  &     $42.18\pm 0.03$  &     $ 21.2\pm  1.2$ & 4, 5, 6, 10  \\
        NGC 5548 &     $ 19.7_{-  1.5}^{+  1.5}$  &     $43.39\pm 0.10$  &     $ 4674\pm   63$  &     $7.92_{-0.04}^{+0.03}$  &     $-1.60_{- 0.49}^{+ 0.46}$  &     $41.79\pm 0.10$  &     $128.1\pm 40.3$  &     $41.63\pm 0.09$  &     $ 89.1\pm 10.0$ & 4, 5, 6, 28  \\
                 &     $ 18.6_{-  2.3}^{+  2.1}$  &     $43.14\pm 0.11$  &     $ 5418\pm  107$  &     $8.03_{-0.06}^{+0.05}$  &     $-1.96_{- 0.51}^{+ 0.47}$  &     $41.61\pm 0.13$  &     $151.3\pm 57.9$  &     $41.63\pm 0.09$  &     $156.9\pm 24.4$ & 4, 5, 6, 28  \\
                 &     $ 15.9_{-  2.5}^{+  2.9}$  &     $43.35\pm 0.09$  &     $ 5236\pm   87$  &     $7.93_{-0.08}^{+0.07}$  &     $-1.65_{- 0.49}^{+ 0.46}$  &     $41.72\pm 0.10$  &     $119.7\pm 37.9$  &     $41.63\pm 0.09$  &     $ 97.3\pm 10.4$ & 4, 5, 6, 28  \\
                 &     $ 11.0_{-  2.0}^{+  1.9}$  &     $43.07\pm 0.11$  &     $ 5986\pm   95$  &     $7.89_{-0.09}^{+0.07}$  &     $-2.07_{- 0.52}^{+ 0.47}$  &     $41.52\pm 0.17$  &     $144.5\pm 66.7$  &     $41.63\pm 0.09$  &     $185.1\pm 31.2$ & 4, 5, 6, 28  \\
                 &     $ 13.0_{-  1.4}^{+  1.6}$  &     $43.32\pm 0.10$  &     $ 5931\pm   42$  &     $7.95_{-0.05}^{+0.05}$  &     $-1.69_{- 0.49}^{+ 0.46}$  &     $41.75\pm 0.09$  &     $135.9\pm 41.2$  &     $41.63\pm 0.09$  &     $103.5\pm 11.3$ & 4, 5, 6, 28  \\
                 &     $ 13.4_{-  4.3}^{+  3.8}$  &     $43.38\pm 0.09$  &     $ 7378\pm   39$  &     $8.15_{-0.17}^{+0.11}$  &     $-1.61_{- 0.49}^{+ 0.46}$  &     $41.73\pm 0.10$  &     $114.4\pm 37.0$  &     $41.63\pm 0.09$  &     $ 91.4\pm  9.8$ & 4, 5, 6, 28  \\
                 &     $ 21.7_{-  2.6}^{+  2.6}$  &     $43.52\pm 0.09$  &     $ 6946\pm   79$  &     $8.31_{-0.06}^{+0.05}$  &     $-1.40_{- 0.48}^{+ 0.45}$  &     $41.82\pm 0.09$  &     $102.4\pm 30.2$  &     $41.63\pm 0.09$  &     $ 65.9\pm  5.1$ & 4, 5, 6, 28  \\
                 &     $ 16.4_{-  1.1}^{+  1.2}$  &     $43.43\pm 0.09$  &     $ 6623\pm   93$  &     $8.15_{-0.03}^{+0.03}$  &     $-1.53_{- 0.48}^{+ 0.45}$  &     $41.75\pm 0.10$  &     $106.3\pm 33.3$  &     $41.63\pm 0.09$  &     $ 80.7\pm  7.7$ & 4, 5, 6, 28  \\
                 &     $ 17.5_{-  1.6}^{+  2.0}$  &     $43.24\pm 0.10$  &     $ 6298\pm   65$  &     $8.13_{-0.04}^{+0.05}$  &     $-1.82_{- 0.49}^{+ 0.46}$  &     $41.72\pm 0.10$  &     $153.5\pm 50.7$  &     $41.63\pm 0.09$  &     $125.8\pm 15.3$ & 4, 5, 6, 28  \\
                 &     $ 26.5_{-  2.2}^{+  4.3}$  &     $43.59\pm 0.09$  &     $ 6177\pm   36$  &     $8.30_{-0.04}^{+0.07}$  &     $-1.30_{- 0.48}^{+ 0.45}$  &     $41.87\pm 0.10$  &     $ 98.1\pm 30.0$  &     $41.63\pm 0.09$  &     $ 56.5\pm  4.8$ & 4, 5, 6, 28  \\
                 &     $ 24.8_{-  3.0}^{+  3.2}$  &     $43.51\pm 0.09$  &     $ 6247\pm   57$  &     $8.28_{-0.06}^{+0.05}$  &     $-1.42_{- 0.48}^{+ 0.45}$  &     $41.83\pm 0.09$  &     $106.6\pm 31.5$  &     $41.63\pm 0.09$  &     $ 68.0\pm  6.0$ & 4, 5, 6, 28  \\
                 &     $  6.5_{-  3.7}^{+  5.7}$  &     $43.11\pm 0.11$  &     $ 6240\pm   77$  &     $7.69_{-0.37}^{+0.27}$  &     $-2.02_{- 0.51}^{+ 0.47}$  &     $41.64\pm 0.13$  &     $172.8\pm 66.1$  &     $41.63\pm 0.09$  &     $170.1\pm 26.8$ & 4, 5, 6, 28  \\
                 &     $ 14.3_{-  7.3}^{+  5.9}$  &     $43.11\pm 0.11$  &     $ 6478\pm  108$  &     $8.07_{-0.31}^{+0.15}$  &     $-2.01_{- 0.51}^{+ 0.47}$  &     $41.55\pm 0.14$  &     $139.8\pm 55.7$  &     $41.63\pm 0.09$  &     $167.5\pm 27.0$ & 4, 5, 6, 28  \\
                 &     $  6.3_{-  2.3}^{+  2.6}$  &     $42.96\pm 0.13$  &     $ 6396\pm  167$  &     $7.70_{-0.20}^{+0.15}$  &     $-2.24_{- 0.55}^{+ 0.49}$  &     $41.12\pm 0.10$  &     $ 74.4\pm 27.3$  &     $41.63\pm 0.09$  &     $240.0\pm 53.5$ & 4, 29        \\
                 &     $  4.2_{-  1.3}^{+  0.9}$  &     $43.01\pm 0.11$  &     $12771\pm   71$  &     $8.12_{-0.16}^{+0.08}$  &     $-2.16_{- 0.51}^{+ 0.47}$  &     $41.33\pm 0.10$  &     $105.3\pm 35.4$  &     $41.63\pm 0.09$  &     $211.4\pm 32.6$ & 4, 17        \\
                 &     $ 12.4_{-  3.9}^{+  2.7}$  &     $42.99\pm 0.11$  &     $11481\pm  574$  &     $8.50_{-0.17}^{+0.09}$  &     $-2.19_{- 0.53}^{+ 0.48}$  &     $41.27\pm 0.10$  &     $ 96.6\pm 34.4$  &     $41.63\pm 0.09$  &     $220.3\pm 40.0$ & 4, 16        \\
                 &     $  7.2_{-  0.3}^{+  1.3}$  &     $43.21\pm 0.09$  &     $ 9912\pm  362$  &     $8.14_{-0.04}^{+0.08}$  &     $-1.87_{- 0.48}^{+ 0.45}$  &     $41.70\pm 0.09$  &     $160.1\pm 46.2$  &     $41.53\pm 0.09$  &     $107.7\pm  7.8$ & 30           \\
                 &     $  4.2_{-  0.4}^{+  0.4}$  &     $43.45\pm 0.09$  &     $ 9496\pm  418$  &     $7.87_{-0.06}^{+0.05}$  &     $-1.51_{- 0.48}^{+ 0.45}$  &     $41.70\pm 0.09$  &     $ 91.2\pm 26.3$  &     $41.55\pm 0.09$  &     $ 64.4\pm  4.5$ & 31           \\
                 & $\bf{ 13.9_{-  6.2}^{+ 11.2}}$ & $\bf{43.30\pm 0.19}$ & $\bf{ 7256\pm 2203}$ & $\bf{8.08_{-0.16}^{+0.16}}$ & $\bf{-1.76_{- 0.32}^{+ 0.31}}$ & $\bf{41.65\pm 0.21}$ & $\bf{117.8\pm 27.3}$ & $\bf{41.58\pm 0.07}$ & $\bf{ 91.3\pm 36.4}$& ...          \\
     PG 1426+015 &     $ 95.0_{- 37.1}^{+ 29.9}$  &     $44.63\pm 0.02$  &     $ 7113\pm  160$  &     $8.97_{-0.22}^{+0.12}$  &     $-1.51_{- 0.28}^{+ 0.47}$  &     $42.83\pm 0.04$  &     $ 80.1\pm  9.2$  &     $42.21\pm 0.03$  &     $ 19.2\pm  1.2$ & 4, 5, 6, 10  \\
         Mrk 817 &     $ 19.0_{-  3.7}^{+  3.9}$  &     $43.79\pm 0.05$  &     $ 4711\pm   49$  &     $7.92_{-0.09}^{+0.08}$  &     $-0.81_{- 0.35}^{+ 0.35}$  &     $42.07\pm 0.05$  &     $ 98.0\pm 16.5$  &     $41.53\pm 0.05$  &     $ 28.2\pm  1.5$ & 4, 5, 6, 7   \\
                 &     $ 15.3_{-  3.5}^{+  3.7}$  &     $43.67\pm 0.05$  &     $ 5237\pm   67$  &     $7.91_{-0.11}^{+0.09}$  &     $-0.98_{- 0.35}^{+ 0.35}$  &     $42.00\pm 0.06$  &     $108.5\pm 20.4$  &     $41.53\pm 0.05$  &     $ 37.0\pm  1.9$ & 4, 5, 6, 7   \\
                 &     $ 33.6_{-  7.6}^{+  6.5}$  &     $43.67\pm 0.05$  &     $ 4767\pm   72$  &     $8.17_{-0.11}^{+0.08}$  &     $-0.98_{- 0.35}^{+ 0.35}$  &     $41.92\pm 0.05$  &     $ 91.1\pm 15.3$  &     $41.53\pm 0.05$  &     $ 36.9\pm  1.9$ & 4, 5, 6, 7   \\
                 &     $ 14.0_{-  3.5}^{+  3.4}$  &     $43.84\pm 0.05$  &     $ 5627\pm   30$  &     $7.94_{-0.12}^{+0.09}$  &     $-0.73_{- 0.35}^{+ 0.35}$  &     $41.77\pm 0.05$  &     $ 43.2\pm  7.1$  &     $41.52\pm 0.05$  &     $ 24.8\pm  1.4$ & 4, 9, 16     \\
                 & $\bf{ 19.9_{-  6.7}^{+  9.9}}$ & $\bf{43.74\pm 0.09}$ & $\bf{ 5348\pm  536}$ & $\bf{7.99_{-0.14}^{+0.14}}$ & $\bf{-0.87_{- 0.22}^{+ 0.22}}$ & $\bf{41.93\pm 0.14}$ & $\bf{ 78.5\pm 34.3}$ & $\bf{41.53\pm 0.04}$ & $\bf{ 31.7\pm  6.4}$& ...          \\
        Mrk 1511 &     $  5.7_{-  0.8}^{+  0.9}$  &     $43.16\pm 0.06$  &     $ 4171\pm  137$  &     $7.29_{-0.07}^{+0.07}$  &     $-0.34_{- 0.24}^{+ 0.24}$  &     $41.52\pm 0.06$  &     $115.5\pm 23.1$  &     $41.02\pm 0.05$  &     $ 36.4\pm  4.1$ & 9, 23        \\
         Mrk 290 &     $  8.7_{-  1.0}^{+  1.2}$  &     $43.17\pm 0.06$  &     $ 4543\pm  227$  &     $7.55_{-0.07}^{+0.07}$  &     $-0.85_{- 0.23}^{+ 0.23}$  &     $41.64\pm 0.06$  &     $153.0\pm 29.0$  &     $41.64\pm 0.06$  &     $150.6\pm 13.2$ & 4, 9, 16     \\
         Mrk 486 &     $ 23.7_{-  2.7}^{+  7.5}$  &     $43.69\pm 0.05$  &     $ 1942\pm   67$  &     $7.24_{-0.06}^{+0.12}$  &     $ 0.55_{- 0.32}^{+ 0.20}$  &     $42.12\pm 0.04$  &     $135.9\pm 20.3$  &     $41.33\pm 0.04$  &     $ 22.3\pm  1.8$ & 1, 2, 3      \\
         Mrk 493 &     $ 11.6_{-  2.6}^{+  1.2}$  &     $43.11\pm 0.08$  &     $  778\pm   12$  &     $6.14_{-0.11}^{+0.04}$  &     $ 1.88_{- 0.21}^{+ 0.33}$  &     $41.35\pm 0.05$  &     $ 87.4\pm 18.1$  &     $40.50\pm 0.05$  &     $ 12.6\pm  1.7$ & 1, 2, 3      \\
     PG 1613+658 &     $ 40.1_{- 15.2}^{+ 15.0}$  &     $44.77\pm 0.02$  &     $ 9074\pm  103$  &     $8.81_{-0.21}^{+0.14}$  &     $-0.97_{- 0.31}^{+ 0.45}$  &     $43.00\pm 0.03$  &     $ 86.7\pm  7.6$  &     $42.59\pm 0.02$  &     $ 33.7\pm  2.2$ & 4, 5, 6, 10  \\
     PG 1617+175 &     $ 71.5_{- 33.7}^{+ 29.6}$  &     $44.39\pm 0.02$  &     $ 6641\pm  190$  &     $8.79_{-0.28}^{+0.15}$  &     $-1.50_{- 0.33}^{+ 0.58}$  &     $42.74\pm 0.05$  &     $114.8\pm 15.1$  &     $41.84\pm 0.03$  &     $ 14.2\pm  0.8$ & 4, 5, 6, 10  \\
     PG 1700+518 &     $251.8_{- 38.8}^{+ 45.9}$  &     $45.59\pm 0.01$  &     $ 2252\pm   85$  &     $8.40_{-0.08}^{+0.08}$  &     $ 1.08_{- 0.17}^{+ 0.17}$  &     $43.78\pm 0.02$  &     $ 78.9\pm  4.5$  &     $42.45\pm 0.02$  &     $  3.7\pm  0.2$ & 4, 5, 6, 10  \\
          3C 382 &     $ 40.5_{-  3.7}^{+  8.0}$  &     $43.84\pm 0.10$  &     $ 7652\pm  383$  &     $8.67_{-0.06}^{+0.09}$  &     $-2.09_{- 0.35}^{+ 0.26}$  &     $42.54\pm 0.03$  &     $259.3\pm 64.0$  &     $41.92\pm 0.03$  &     $ 61.4\pm 13.9$ & 12           \\
        3C 390.3 &     $ 23.6_{-  6.7}^{+  6.2}$  &     $43.68\pm 0.10$  &     $12694\pm   13$  &     $8.87_{-0.15}^{+0.10}$  &     $-3.35_{- 0.65}^{+ 0.60}$  &     $42.29\pm 0.05$  &     $206.2\pm 50.7$  &     $42.20\pm 0.03$  &     $169.1\pm 36.9$ & 4, 5, 6, 30  \\
                 &     $ 46.4_{-  3.2}^{+  3.6}$  &     $44.50\pm 0.03$  &     $13211\pm   28$  &     $9.20_{-0.03}^{+0.03}$  &     $-2.12_{- 0.51}^{+ 0.51}$  &     $42.78\pm 0.04$  &     $ 97.1\pm 10.0$  &     $42.27\pm 0.03$  &     $ 30.5\pm  0.8$ & 4, 31        \\
                 & $\bf{ 44.5_{- 17.0}^{+ 27.6}}$ & $\bf{44.43\pm 0.58}$ & $\bf{12796\pm  361}$ & $\bf{9.18_{-0.23}^{+0.23}}$ & $\bf{-2.62_{- 0.96}^{+ 0.95}}$ & $\bf{42.60\pm 0.35}$ & $\bf{108.8\pm 58.8}$ & $\bf{42.24\pm 0.05}$ & $\bf{ 31.2\pm 37.8}$& ...          \\
    KA 1858+4850 &     $ 13.5_{-  2.3}^{+  2.0}$  &     $43.43\pm 0.05$  &     $ 1820\pm   79$  &     $6.94_{-0.09}^{+0.07}$  &     $ 0.75_{- 0.21}^{+ 0.25}$  &     $41.89\pm 0.04$  &     $146.9\pm 21.1$  &     $41.40\pm 0.03$  &     $ 47.8\pm  5.3$ & 32           \\
        NGC 6814 &     $  6.6_{-  0.9}^{+  0.9}$  &     $42.12\pm 0.28$  &     $ 3323\pm    7$  &     $7.16_{-0.06}^{+0.05}$  &     $-1.64_{- 0.80}^{+ 0.46}$  &     $40.50\pm 0.28$  &     $121.6\pm112.2$  &     $40.26\pm 0.28$  &     $ 69.8\pm 10.0$ & 4, 17        \\
         Mrk 509 &     $ 79.6_{-  5.4}^{+  6.1}$  &     $44.19\pm 0.05$  &     $ 3015\pm    2$  &     $8.15_{-0.03}^{+0.03}$  &     $-0.52_{- 0.14}^{+ 0.13}$  &     $42.61\pm 0.04$  &     $132.7\pm 19.1$  &     $42.38\pm 0.05$  &     $ 77.7\pm  5.2$ & 4, 5, 6, 7   \\
     PG 2130+099 &     $ 22.6_{-  3.6}^{+  2.7}$  &     $44.32\pm 0.04$  &     $ 2101\pm  100$  &     $7.29_{-0.09}^{+0.06}$  &     $ 1.40_{- 0.19}^{+ 0.24}$  &     $42.77\pm 0.04$  &     $142.4\pm 18.4$  &     $42.20\pm 0.04$  &     $ 38.4\pm  4.3$ & 4, 8, 9      \\
        NGC 7469 &     $ 10.8_{-  1.3}^{+  3.4}$  &     $43.51\pm 0.11$  &     $ 4369\pm    6$  &     $7.60_{-0.06}^{+0.12}$  &     $-0.46_{- 0.42}^{+ 0.26}$  &     $41.60\pm 0.10$  &     $ 63.0\pm 21.0$  &     $41.69\pm 0.09$  &     $ 77.4\pm 10.8$ & 4, 33        \\
\enddata
\tablecomments{The objects are sorted in the order of right ascension. References: (1) \cite{du2014}, (2) \cite{wang2014}, (3) \cite{hu2015}, 
(4) \cite{bentz2013}, (5) \cite{collin2006}, (6) \cite{kaspi2005}, (7) \cite{peterson1998}, 
(8) \cite{grier2012}, (9) \cite{du2015}, (10) \cite{kaspi2000}, (11) \cite{santos_lleo1997}, 
(12) \cite{fausnaugh2017}, (13) \cite{du2018}, (14) \cite{du2016V}, (15) \cite{bentz2009b}, 
(16) \cite{denney2010}, (17) \cite{bentz2009}, (18) \cite{stirpe1994}, (19) \cite{bentz2016a}, 
(20) \cite{bentz2006}, (21) \cite{zhang2019}, (22) \cite{denney2006}, (23) \cite{barth2013}, 
(24) \cite{bentz2016b}, (25) \cite{hu2016}, (26) \cite{bentz2014}, (27) \cite{santos_lleo2001}, 
(28) \cite{peterson2002}, (29) \cite{bentz2007}, (30) \cite{lu2016}, (31) \cite{pei2017}, 
(32) \cite{dietrich1998}, (33) \cite{dietrich2012}, (34) \cite{pei2014}, 
(35) \cite{peterson2014}. For the objects with multiple observations, the \dotm\ of 
the individual campaigns are calculated based on the averaged \mbh.}
\end{deluxetable*}

\startlongtable
\begin{deluxetable*}{lcccccc}
\tablecolumns{7}
\tablecaption{Single-epoch spectral properties\label{tab:properties}}
\tabletypesize{\scriptsize}
\tablehead{
    \colhead{Objects}                             &
    \colhead{${\cal R}_{\rm Fe}$}                 &
    \colhead{${\cal D}_{\hb}$}                    &
    \colhead{${\rm FWHM_{Fe}/FWHM_{\hb}}$}        &
    \colhead{Asymmetry}                           &
    \colhead{EW(\heii)}                           &
    \colhead{Ref.}                                \\ \cline{2-7}
    \colhead{}                                    &
    \colhead{}                                    &
    \colhead{}                                    &
    \colhead{}                                    &
    \colhead{}                                    &
    \colhead{(\AA)}                               &
    \colhead{}
}
\startdata
         Mrk 335 & $0.62$ & $1.27\pm0.05$ & $0.93\pm0.10$ & $-0.025\pm 0.007$ & $23.9\pm 0.3$ & 1, 2, 3$^a$          \\ 
     PG 0026+129 & $0.33$ & $1.46\pm0.09$ & $0.67\pm0.02$ & $ 0.036\pm 0.006$ & $13.6\pm 0.2$ & 1, 4, 5$^a$          \\ 
     PG 0052+251 & $0.12$ & $2.31\pm0.05$ & $0.51\pm0.02$ & $-0.037\pm 0.004$ & $28.4\pm 1.2$ & 1, 4, 6$^a$          \\ 
        Fairall9 & $0.49$ & $2.56\pm0.03$ & $0.60\pm0.01$ & $-0.233\pm 0.022$ & $12.1\pm 0.3$ & 1, 8, 9$^a$          \\ 
         Mrk 590 & $0.45$ & $1.39\pm0.07$ & $0.58\pm0.02$ & $-0.025\pm 0.010$ & $47.1\pm 2.7$ & 1, 10$^a$            \\ 
        Mrk 1044 & $0.99$ & $1.54\pm0.03$ & $0.74\pm0.03$ & $ 0.037\pm 0.003$ & $23.0\pm 0.9$ & 1, 2, 11$^a$         \\ 
          3C 120 & $0.39$ & $1.86\pm0.05$ & $0.54\pm0.12$ & $-0.187\pm 0.020$ & $17.6\pm 1.6$ & 1, 12$^a$            \\ 
 IRAS 04416+1215 & $1.96$ & $1.44\pm0.06$ & $0.86\pm0.04$ & $ 0.195\pm 0.012$ & too weak      & 1, 2, 11$^a$, 13$^a$ \\ 
         Ark 120 & $0.83$ & $1.65\pm0.01$ & $0.57\pm0.11$ & $-0.185\pm 0.020$ & too weak      & 1, 12$^a$, 14$^a$    \\ 
  MCG +08-11-011 & $0.29$ & $1.55\pm0.11$ & $0.56\pm0.10$ & $-0.293\pm 0.020$ & $14.6\pm 0.7$ & 15$^a$               \\ 
         Mrk 374 & $0.88$ & $1.38\pm0.10$ & $0.32\pm0.07$ & $-0.100\pm 0.016$ & $ 5.8\pm 0.4$ & 15$^a$               \\ 
          Mrk 79 & $0.33$ & $2.10\pm0.06$ & $0.38\pm0.18$ & $-0.107\pm 0.018$ & $22.8\pm 0.8$ & 1, 16$^a$            \\ 
    SDSS J074352 & $1.11$ & $1.60\pm0.02$ & $0.79\pm0.02$ & $-0.019\pm 0.003$ & too weak      & 11$^a$, 17$^a$       \\ 
    SDSS J075051 & $1.22$ & $1.54\pm0.01$ & $1.04\pm0.06$ & $ 0.076\pm 0.018$ & $ 5.7\pm 0.7$ & 11$^a$, 17$^a$       \\ 
    SDSS J075101 & $0.97$ & $1.52\pm0.08$ & $1.27\pm0.10$ & $ 0.042\pm 0.020$ & $ 5.7\pm 0.4$ & 1, 11$^a$            \\ 
         Mrk 382 & $0.75$ & $1.74\pm0.36$ & $0.91\pm0.24$ & $ 0.112\pm 0.019$ & $15.1\pm 0.4$ & 1, 2, 11$^a$, 13     \\ 
    SDSS J075949 & $1.02$ & $1.49\pm0.04$ & $0.91\pm0.03$ & $ 0.067\pm 0.011$ & $15.4\pm 1.0$ & 11$^a$, 17$^a$       \\ 
    SDSS J080101 & $1.01$ & $1.61\pm0.08$ & $0.88\pm0.02$ & $ 0.040\pm 0.005$ & $ 6.9\pm 0.4$ & 1, 11$^a$, 18        \\ 
    SDSS J080131 & $1.49$ & $1.36\pm0.03$ & $1.10\pm0.07$ & $ 0.059\pm 0.012$ & $ 5.1\pm 1.3$ & 1, 11$^a$, 19        \\ 
     PG 0804+761 & $0.61$ & $2.13\pm0.04$ & $0.53\pm0.01$ & $-0.007\pm 0.001$ & $ 1.9\pm 0.1$ & 4, 5$^a$, 6          \\ 
    SDSS J081441 & $0.46$ & $1.54\pm0.08$ & $0.71\pm0.02$ & $-0.008\pm 0.003$ & $20.3\pm 0.8$ & 1, 11$^a$, 17        \\ 
    SDSS J081456 & $1.31$ & $1.71\pm0.09$ & $0.77\pm0.01$ & $ 0.070\pm 0.007$ & $ 8.5\pm 0.7$ & 1, 11$^a$, 18        \\ 
        NGC 2617 & $0.31$ & $2.55\pm0.18$ & $0.22\pm0.04$ & $-0.169\pm 0.020$ & too weak      & 12, 16$^a$, 20       \\ 
    SDSS J083553 & $1.57$ & $1.73\pm0.02$ & $0.87\pm0.02$ & $ 0.050\pm 0.006$ & $12.4\pm 0.8$ & 11, 17$^a$           \\ 
    SDSS J084533 & $1.11$ & $1.38\pm0.03$ & $0.98\pm0.03$ & $ 0.083\pm 0.017$ & $ 5.5\pm 0.8$ & 1, 11$^a$, 19        \\ 
     PG 0844+349 & $0.78$ & $1.79\pm0.04$ & $0.82\pm0.02$ & $ 0.071\pm 0.009$ & $ 9.4\pm 0.8$ & 1, 4, 6, 21$^a$      \\ 
    SDSS J085946 & $1.39$ & $1.59\pm0.04$ & $0.92\pm0.03$ & $ 0.091\pm 0.020$ & $ 5.2\pm 1.4$ & 1, 11$^a$, 19        \\ 
         Mrk 110 & $0.14$ & $1.69\pm0.09$ & $0.63\pm0.08$ & $-0.032\pm 0.005$ & $21.0\pm 0.6$ & 1, 21$^a$            \\ 
    SDSS J093302 & $1.44$ & $1.26\pm0.02$ & $1.09\pm0.04$ & $ 0.052\pm 0.005$ & $16.3\pm 0.6$ & 11$^a$, 17           \\ 
    SDSS J093922 & $1.48$ & $1.29\pm0.06$ & $1.17\pm0.05$ & $-0.051\pm 0.021$ & $16.6\pm 1.0$ & 1, 11$^a$, 18        \\ 
     PG 0953+414 & $0.27$ & $1.85\pm0.04$ & $0.50\pm0.02$ & $-0.009\pm 0.004$ & $ 7.1\pm 1.8$ & 1, 4, 6, 21$^a$      \\ 
    SDSS J100402 & $1.17$ & $1.47\pm0.01$ & $0.81\pm0.02$ & $ 0.099\pm 0.007$ & $ 0.9\pm 0.3$ & 11$^a$, 17           \\ 
    SDSS J101000 & $2.17$ & $1.64\pm0.00$ & $0.84\pm0.03$ & $ 0.104\pm 0.015$ & $ 4.8\pm 0.8$ & 11$^a$, 17           \\ 
        NGC 3227 & $0.46$ & $2.44\pm0.17$ & $0.71\pm0.05$ & $-0.011\pm 0.002$ & $33.0\pm 1.5$ & 1, 7, 9$^a$, 22$^a$  \\ 
    SDSS J102339 & $1.03$ & $1.48\pm0.04$ & $0.66\pm0.05$ & $-0.007\pm 0.012$ & $20.5\pm 2.9$ & 1, 11$^a$, 19        \\ 
         Mrk 142 & $1.14$ & $1.36\pm0.26$ & $0.95\pm0.06$ & $ 0.077\pm 0.005$ & $14.8\pm 0.4$ & 1, 11$^a$            \\ 
        NGC 3516 & $0.66$ & $2.45\pm0.17$ & $0.57\pm0.02$ & $-0.034\pm 0.004$ & $47.7\pm 4.8$ & 1, 7, 9$^a$          \\ 
   SBS 1116+583A & $0.59$ & $2.36\pm0.13$ & $0.66\pm0.04$ & $-0.029\pm 0.008$ & $31.2\pm 1.1$ & 1, 11$^a$, 23        \\ 
         Arp 151 & $0.32$ & $1.54\pm0.04$ & $0.84\pm0.03$ & $-0.174\pm 0.008$ & $41.3\pm 2.3$ & 1, 23$^a$            \\ 
        NGC 3783 & $0.04$ & $2.23\pm0.05$ & $0.51\pm0.04$ & $-0.088\pm 0.011$ & $57.9\pm 2.5$ & 1, 4, 24, 25$^a$     \\ 
  MCG +06-26-012 & $1.04$ & $1.70\pm0.11$ & $0.87\pm0.07$ & $ 0.026\pm 0.001$ & $16.7\pm 0.5$ & 1, 2, 13$^a$         \\ 
       UGC 06728 & $1.11$ & $0.89\pm0.12$ & $0.60\pm0.12$ & $-0.069\pm 0.020$ & $14.8\pm 2.9$ & 26$^a$               \\ 
        Mrk 1310 & $0.46$ & $1.99\pm0.07$ & $0.68\pm0.04$ & $-0.005\pm 0.002$ & $32.2\pm 5.2$ & 1, 23$^a$            \\ 
        NGC 4051 & $1.18$ & $1.97\pm0.15$ & $1.36\pm0.12$ & $ 0.018\pm 0.009$ & $ 8.9\pm 0.6$ & 1, 7, 27, 28$^a$     \\ 
        NGC 4151 & $0.22$ & $2.76\pm0.07$ & $0.34\pm0.02$ & $ 0.010\pm 0.017$ & $21.5\pm 1.6$ & 1, 4, 9$^a$, 29      \\ 
     PG 1211+143 & $0.42$ & $1.35\pm0.04$ & $0.88\pm0.03$ & $ 0.007\pm 0.007$ & $15.2\pm 0.6$ & 1, 4, 6, 21$^a$      \\ 
         Mrk 202 & $0.57$ & $1.70\pm0.08$ & $0.73\pm0.22$ & $-0.015\pm 0.006$ & $21.1\pm 0.7$ & 1, 11$^a$, 23        \\ 
        NGC 4253 & $0.99$ & $1.48\pm0.06$ & $0.72\pm0.03$ & $-0.010\pm 0.001$ & $29.7\pm 1.3$ & 1, 23$^a$            \\ 
     PG 1226+023 & $0.64$ & $1.97\pm0.03$ & $0.64\pm0.02$ & $-0.079\pm 0.004$ & $ 0.7\pm 0.1$ & 30$^a$               \\ 
     PG 1229+204 & $0.53$ & $2.38\pm0.05$ & $0.80\pm0.02$ & $-0.018\pm 0.007$ & $13.6\pm 1.1$ & 1, 4, 6, 21$^a$      \\ 
        NGC 4593 & $0.89$ & $2.87\pm0.66$ & $0.76\pm0.07$ & $ 0.043\pm 0.020$ & $ 6.2\pm 0.2$ & 1, 31$^a$            \\ 
IRAS F12397+3333 & $1.48$ & $1.57\pm0.52$ & $0.97\pm0.30$ & $ 0.084\pm 0.010$ & $19.6\pm 0.6$ & 1, 2, 3, 11$^a$      \\ 
        NGC 4748 & $0.99$ & $1.93\pm0.08$ & $0.91\pm0.05$ & $ 0.011\pm 0.002$ & $34.0\pm 1.3$ & 1, 23$^a$            \\ 
     PG 1307+085 & $0.21$ & $2.58\pm0.09$ & $0.40\pm0.06$ & $-0.207\pm 0.006$ & $15.7\pm 0.7$ & 1, 4, 6$^a$          \\ 
  MCG +06-30-015 & $0.93$ & $2.01\pm0.08$ & $1.01\pm0.49$ & $ 0.007\pm 0.020$ & $16.6\pm 7.3$ & 23$^a$, 32           \\ 
        NGC 5273 & $0.58$ & $3.12\pm0.13$ & $0.71\pm0.04$ & $ 0.022\pm 0.005$ & too weak      & 1, 22$^a$, 33        \\ 
         Mrk 279 & $0.55$ & $2.94\pm0.03$ & $0.52\pm0.10$ & $ 0.061\pm 0.006$ & $ 7.9\pm 0.2$ & 1, 4, 34, 35$^a$     \\ 
     PG 1411+442 & $0.63$ & $1.58\pm0.04$ & $0.61\pm0.04$ & $-0.040\pm 0.004$ & $11.9\pm 0.2$ & 1, 4, 5$^a$, 6       \\ 
        NGC 5548 & $0.10$ & $2.66\pm0.33$ & $0.46\pm0.04$ & $-0.193\pm 0.012$ & $ 8.4\pm 0.8$ & 1, 4, 11$^a$, 36     \\ 
     PG 1426+015 & $0.46$ & $2.45\pm0.09$ & $0.72\pm0.01$ & $-0.140\pm 0.003$ & $10.4\pm 0.6$ & 1, 4, 6$^a$          \\ 
         Mrk 817 & $0.69$ & $2.59\pm0.29$ & $0.87\pm0.07$ & $-0.163\pm 0.057$ & $22.7\pm 1.1$ & 1, 4, 37, 38$^a$     \\ 
        Mrk 1511 & $0.80$ & $2.20\pm0.13$ & $0.75\pm0.04$ & $-0.003\pm 0.020$ & $ 6.3\pm 0.2$ & 1, 31$^a$            \\ 
         Mrk 290 & $0.29$ & $2.57\pm0.18$ & $0.73\pm0.03$ & $-0.019\pm 0.004$ & $21.2\pm 1.1$ & 1, 7, 21, 23$^a$     \\ 
         Mrk 486 & $0.54$ & $1.50\pm0.06$ & $0.92\pm0.06$ & $ 0.035\pm 0.003$ & $15.7\pm 0.2$ & 1, 2, 11$^a$, 13$^a$ \\ 
         Mrk 493 & $1.13$ & $1.52\pm0.03$ & $1.00\pm0.02$ & $ 0.025\pm 0.018$ & $ 5.8\pm 0.3$ & 1, 2, 11$^a$, 13$^a$ \\ 
     PG 1613+658 & $0.38$ & $2.94\pm0.05$ & $0.35\pm0.02$ & $-0.167\pm 0.003$ & $ 7.6\pm 0.4$ & 1, 4, 6$^a$, 21      \\ 
     PG 1617+175 & $0.74$ & $2.87\pm0.12$ & $0.98\pm0.04$ & $-0.068\pm 0.020$ & $ 1.1\pm 0.2$ & 1, 4, 6$^a$          \\ 
     PG 1700+518 & $1.32$ & $1.09\pm0.08$ & $0.69\pm0.06$ & $ 0.344\pm 0.024$ & too weak      & 1, 4, 5, 6, 27, 39   \\ 
          3C 382 & $0.31$ & $1.87\pm0.13$ & $0.71\pm0.13$ & $-0.182\pm 0.011$ & too weak      & 15, 35$^a$           \\ 
        3C 390.3 & $0.12$ & $2.62\pm0.66$ & $0.47\pm0.08$ & $-0.062\pm 0.020$ & too weak      & 1, 18, 40, 41        \\ 
    KA 1858+4850 & $0.11$ & $2.13\pm0.13$ & $0.46\pm0.05$ & $ 0.015\pm 0.020$ & $35.2\pm 9.2$ & 1, 42$^a$            \\ 
        NGC 6814 & $0.45$ & $1.73\pm0.03$ & $0.73\pm0.01$ & $-0.010\pm 0.020$ & $23.8\pm 0.3$ & 1, 18, 23$^a$        \\ 
         Mrk 509 & $0.13$ & $1.94\pm0.01$ & $0.59\pm0.24$ & $-0.057\pm 0.022$ & $44.0\pm10.0$ & 1, 4, 9$^a$, 37      \\ 
     PG 2130+099 & $0.96$ & $1.39\pm0.11$ & $0.95\pm0.06$ & $ 0.017\pm 0.001$ & $14.2\pm 0.2$ & 1, 5$^a$, 43         \\ 
        NGC 7469 & $0.43$ & $1.49\pm0.18$ & $0.83\pm0.24$ & $-0.316\pm 0.032$ & $17.5\pm 0.7$ & 1, 27, 44, 45$^a$
\enddata
\tablecomments{References: (1) \cite{du2016FP}, (2) \cite{hu2015}, (3) \cite{du2014}, 
(4) \cite{collin2006}, (5) \cite{hu2019}, (6) \cite{kaspi2000}, (7) \cite{denney2010}, 
(8) \cite{santos_lleo1997}, (9) the spectrum from the archive of Hubble Space Telescope, 
(10) \cite{marziani2003}, (11) the spectrum from the SDSS archive,
(12) \cite{du2018b}, (13) \cite{wang2014}, (14) \cite{doroshenko2008}, 
(15) \cite{fausnaugh2017}, (16) \cite{brotherton2019}, (17) \cite{du2018}, 
(18) \cite{du2015}, (19) \cite{du2016V}, (20) \cite{fausnaugh2017}, (21) \cite{boroson1992}, 
(22) \cite{ho1995}, (23) \cite{bentz2009}, (24) \cite{stirpe1994}, (25) \cite{jones2009}, 
(26) \cite{bentz2016a}, (27) \cite{kollatschny2011}, (28) \cite{moustakas2006}, 
(29) \cite{bentz2006}, (30) \cite{zhang2019}, (31) \cite{barth2013}, 
(32) \cite{hu2016}, (33) \cite{bentz2014}, (34) \cite{santos_lleo2001}, 
(35) \cite{marziani2003}, (36) \cite{peterson2002}, (37) \cite{peterson1998}, 
(38) \cite{ilic2006}, (39) \cite{bian2010}, (40) \cite{dietrich2012}, (41) \cite{du2018b}, 
(42) \cite{pei2014}, (43) \cite{grier2012}, (44) \cite{peterson2014}, 
(45) \cite{kim1995}.
$^a$ The spectral properties are measured from the spectrum in the reference if 
not found in the literatures. As in \cite{du2016FP}, we adopt 20\% of the \Rfe\ as 
its error bar.
}
\end{deluxetable*}

\end{document}